\PassOptionsToPackage{dvipsnames}{xcolor}

\documentclass[acmsmall,screen,nonacm]{acmart}

\usepackage[normalem]{ulem}

\AtBeginDocument{%
  }

\newif\ifshowcomments
\showcommentsfalse

\ifshowcomments
  \newcommand{\evan}[1]{{\color{blue}\textbf{evan:} #1}}
  \newcommand{\fixme}[1]{{\color{red}\textbf{fix me:} #1}}
  \newcommand{\todo}[1]{{\color{red}\textbf{TODO:} #1}}
  
  \newcommand{\justin}[1]{{\color{magenta}\textbf{justin:} #1}}
  \newcommand{\chasins}[1]{{\color{cyan}\textbf{sarah:} #1}}
  \newcommand{\owh}[1]{{\color{olive}\textbf{owh:} #1}}
  \newcommand{\rl}[1]{{\color{orange}\textbf{rubens:} #1}}
\else
  \newcommand{\evan}[1]{}
  \newcommand{\fixme}[1]{}
  \newcommand{\todo}[1]{}
    
  \newcommand{\justin}[1]{}
  \newcommand{\chasins}[1]{}
  \newcommand{\owh}[1]{}
  \newcommand{\rl}[1]{}
\fi

\newcommand{\name}{\textbf{\textsc{Voya}}}

\AtEndPreamble{%
    \theoremstyle{acmdefinition}
    \newtheorem{remark}[theorem]{Remark}}

\usepackage[normalem]{ulem}
\usepackage{booktabs}
\usepackage{tabularx}
\usepackage{multirow}
\usepackage{colortbl}
\usepackage{xspace}
\usepackage{microtype}
\usepackage{calc}        
\usepackage{mathtools}
\usepackage{amsmath}
\usepackage{amsfonts}
\usepackage{mathpartir}  
\usepackage{float}
\usepackage{wrapfig}
\usepackage{enumitem}
\usepackage{xcolor,xcolor-solarized}
\usepackage{listings}
\usepackage{tikz}
\usepackage[edges]{forest}
\usetikzlibrary{positioning,fit,arrows}
\usepackage{algorithm}
\usepackage[noend]{algpseudocode}
\usepackage{cleveref}
\usepackage{subcaption}

\lstdefinestyle{base}{
    xleftmargin=2em,
    columns=fullflexible,
    numbers=left,
    numbersep=5pt,
    keywordstyle=\color{solarized-blue},
    commentstyle=\color{solarized-base01},
    stringstyle=\color{solarized-magenta},
    numberstyle=\color{solarized-base02},
    emphstyle=[1]\color{solarized-red},
    emphstyle=[2]\color{solarized-green},
    emphstyle=[3]\color{solarized-magenta},
    basicstyle={\footnotesize\linespread{0.8}\color{solarized-base03}\ttfamily},
    escapeinside={(*}{*)}
  }
\lstdefinestyle{tight}{
    basicstyle={\footnotesize\linespread{0.8}\color{solarized-base03}\ttfamily},
  }

\lstdefinestyle{menu}{
  style=base,
  comment=[l]{//},
  morekeywords={compute_root,compute_at,store_at,inline,build_pipeline,
                order,serial,split,fuse,reorder,vectorize,parallel},
  emph=[2]{li2018,mullapudi2016,adams2019,taco,scorch,hand,Li2018,Mullapudi2016,Adams2019,TACO,Scorch,Hand},
  emph=[3]{dense,csr,csc,dcsr,coo,custom},
}

\lstdefinestyle{smt}{
  style=base,
  numbers=none,
  comment=[l]{;},
  morekeywords={declare-const,assert,check-sat,Int},
  emph=[1]{unsat,sat},
}

\lstdefinestyle{halide}{
  style=base,
  comment=[l]{//},
  morekeywords={compute_root,compute_at,store_at,compute_inline,store_in,
                split,fuse,reorder,vectorize,parallel,unroll,serial,tile,bound},
  emph=[2]{GuardWithIf,TailStrategy,MemoryType,Stack},  
}

\lstdefinestyle{halide-algorithm}{
  style=halide,
  emph=[3]{gray,blur_y,blur_x,sharpen,ratio,output,
           kernel,input,input_bounded},
}

\newcommand\code[1]{\lstinline[mathescape=true,basicstyle=\ttfamily\normalsize]|#1|}

\algnewcommand{\LineComment}[1]{\State // #1}

\colorlet{hlChan}{solarized-yellow!35}   
\colorlet{hlPlace}{solarized-cyan!25}    
\newcommand{\hlbox}[2]{{\setlength{\fboxsep}{0.6pt}\colorbox{#1}{\vphantom{Ag}\ttfamily #2}}}
\newcommand{\hlChan}[1]{\hlbox{hlChan}{#1}}
\newcommand{\hlPlace}[1]{\hlbox{hlPlace}{#1}}
\DeclareRobustCommand{\chg}[1]{\tikz[baseline=(c.base)]\node[circle,draw=solarized-base01,inner sep=0.5pt,font=\tiny\sffamily](c){#1};}

\newcommand{\CStrongCompleteness}{\textbf{\textsc{Strong Completeness}}}
\newcommand{\CStrongSoundness}{\textbf{\textsc{Strong Soundness}}}

\newcommand{\Sis}{::=}
\newcommand{\Sor}{\mathrel{\makebox[\widthof{$\Sis$}]{$|$}}}

\definecolor{pathgreen}{HTML}{3D8B74}

\newcommand{\HName}{h}
\newcommand{\Fun}{f}
\newcommand{\Exp}{{\color{RawSienna}e}}
\newcommand{\ExpSub}[1]{{\color{RawSienna}e_{#1}}}

\newcommand{\Start}{\ExpSub{\textsf{start}}}

\newcommand{\Hole}[1]{{?_{#1}}}
\newcommand{\App}[2]{#1(#2)}
\newcommand{\Step}{{\color{Violet}\sigma}}
\newcommand{\StepSub}[1]{{\color{Violet}\sigma_{#1}}}
\newcommand{\StepSet}{{\color{Violet}\Sigma}}
\newcommand{\StepSetSub}[1]{{\color{Violet}\Sigma_{#1}}}
\newcommand{\StepTD}[2]{{\color{Violet} \Hole{#1} \rightsquigarrow {#2}}}

\newcommand{\StepSeq}[2]{{\color{Violet} #1 \, ;\, #2}}

\newcommand{\ExpList}{\ExpSub{1}, \ldots, \ExpSub{N}}

\newcommand{\StepD}[3]{#1 \xrightarrow{#2} #3}

\newcommand{\Alg}{A}
\newcommand{\Src}{\mathcal{S}}
\newcommand{\Tgt}{\mathcal{T}}
\newcommand{\ValidA}[1]{#1\, {\color{PineGreen}\textsf{valid}_{\Alg}}}
\newcommand{\TraceSet}{{\color{Violet}\Theta}}
\newcommand{\HoleList}[1]{\Hole{#1_1}, \ldots, \Hole{#1_N}}
\newcommand{\Hint}{\eta}
\newcommand{\Ingest}[3]{#1 \vdash #2 \Downarrow #3}
\newcommand{\TraceSetSub}[1]{{\color{Violet}\Theta_{#1}}}

\newcommand{\Denot}[1]{\llbracket #1 \rrbracket}

\newcommand{\ValStr}[1]{\mathfrak{V}}
\newcommand{\EveSig}[1]{\mathbb{D}}

\begin{document}

\title{Interactive Compiler Scheduling by Navigation}

\author{Evan Williams}
\email{evanwilliams@cmu.edu}
\orcid{0009-0006-5113-7105}
\affiliation{
    \institution{Carnegie Mellon University}
    \city{Pittsburgh}
    \state{Pennsylvania}
    \country{USA}
}

\author{Justin Lubin}
\email{justinlubin@berkeley.edu}
\orcid{0000-0003-2311-1873}
\affiliation{
    \institution{University of California, Berkeley}
    \city{Berkeley}
    \state{California}
    \country{USA}
}

\author{Rubens Lacouture}
\email{rubensl@stanford.edu}
\orcid{0009-0008-2268-0074}
\affiliation{
    \institution{Stanford University}
    \city{Stanford}
    \state{California}
    \country{USA}
}

\author{Sarah E. Chasins}
\email{schasins@berkeley.edu}
\orcid{0000-0003-0557-3580}
\affiliation{
    \institution{University of California, Berkeley}
    \city{Berkeley}
    \state{California}
    \country{USA}
}

\author{Olivia Hsu}
\email{owh@cmu.edu}
\orcid{0000-0002-4195-8106}
\affiliation{
    \institution{Carnegie Mellon University}
    \city{Pittsburgh}
    \state{Pennsylvania}
    \country{USA}
}

\renewcommand{\shortauthors}{Williams et al.}

\begin{abstract}
Compiler scheduling languages enable productive programming by separating high-level algorithms from schedules that determine how they are optimized.
In current domain-specific compilers, programmers mainly choose between manual scheduling, which can reach any schedule but demands complex reasoning, and autoscheduling, which often restricts the scheduling space and falls short of expert performance.
These approaches impose a tradeoff between performance and productivity, whereas programmers ideally want the benefits of both.
We present \name{}, an interactive synthesis tool that enables programmers to compose manual and autoscheduling approaches, with strong validity and reachability guarantees for both final schedules and the interactive process.
\name{} lets programmers build schedules one decision at a time, using existing handwritten and autoscheduler-generated schedules as sources of hints.
We instantiate \name{} for two domain-specific compilers: Halide on CPUs and FuseFlow for sparse deep-learning models on dataflow accelerators.
Schedules authored with \name{} can outperform the fastest existing expert-written or autoscheduler-generated schedules by up to 1.8$\times$ in Halide and 7.7$\times$ in FuseFlow.
LLM agents using \name{} match the performance of agents writing FuseFlow schedules directly with 2.3$\times$ fewer output tokens on average. 
Our results suggest that any compiler with a scheduling language can support interactive scheduling with strong guarantees.
\end{abstract}

\maketitle

\section{Introduction}
\label{sec:intro}

Domain-specific languages (DSLs), and their optimizing compilers, are pervasive in high-performance computing environments. 
For example, Halide~\cite{ragan-kelley:halide} powers Adobe Photoshop's compositing engine for pixel blending operations~\cite{adobe:halide}, and TVM~\cite{chen:tvm} is used for production deep-learning models for CPUs, GPUs, and specialized accelerators~\cite{apache:tvm}. 
These languages separate the \emph{algorithm} (what to compute) from its \emph{schedule} (how to compute it)~\cite{ragan-kelley:decouple, ragan-kelley:halide, baghdadi:tiramisu, ikarashi:exo, ikarashi:exo2, koeplinger:spatial, lai:hetero-cl, chen:tvm, kjolstad:taco-sched, lacouture:fuseflow, hsu:stardust, hsu:sam, hall:taxonomy, gyurgyik:scion, chen:allo, senanayake:sparse-iteration, hu:taichi, yadav:distal, yadav:spdistal, sundram:recuma, tariq:reptile, bansal:mosaic, hagedorn:rise-elevate, nayak:teaal, schulze:mdh, zhang:graphit, hagedorn:fireiron, fegade:cora, venkat:swirl, fegade:cortex}.
The design of the scheduling language and the right choice of schedule enable the high-level, domain-specific algorithm to be implemented with high performance. 
A carefully crafted Halide schedule for unsharp masking, a common image-sharpening filter (\Cref{sec:motivating}), yields a speedup of 273$\times$ over the default schedule.

Although the best schedules are typically written by hand, discovering them requires deep expertise in performance optimization.
The programmer must draw on their knowledge of hardware and input characteristics to decide which scheduling directive, if any, to apply next. 
Scheduling directives also depend on one another, making manual scheduling even more difficult.
As a result, a programmer cannot easily predict whether an edit yields a valid schedule, i.e., one that compiles and runs successfully. 
The space of valid schedules is vast: unsharp masking alone has over $10^{20}$, even when restricted to a subset of Halide's directives.
Schedule authors often start from a simple schedule they understand and iteratively make local edits to it.

The complexity of writing schedules has motivated a substantial body of work on \emph{autoscheduling}~\cite{mullapudi:halide-auto, mullapudi:polymage, adams:learning, anderson:halide-gpu, ahrens:auto, won:waco, dias:sparseauto, yan:scorch, hellsten:baco, zheng:ansor, nardi:hyper-mapper, ansel:opentuner, zhao:nautilus, shao:metaschedule, zheng:amos, chen:autotvm, zheng:flextensor, haj-ali:protuner, baghdadi:tiramisu-auto, sioutas:halide-gpu, sioutas:reuse}.
While autoschedulers eliminate the need for manual scheduling, their schedules often fall short of expert-written ones; for unsharp masking, the handwritten schedule achieves a 1.3--9.7$\times$ speedup over all three CPU autoschedulers included in the Halide distribution~\cite{mullapudi:halide-auto, li:differentiable-halide, adams:learning}.
Autoschedulers search a fixed subset of the scheduling space and can miss better schedules when they fall outside that subset. 
This is especially apparent for data-dependent workloads, such as sparse tensor algebra~\cite{hsu:sam, won:waco, bansal:mosaic, swann:seer, venkat:loop, kjolstad:taco-sched}.
Autoschedulers also offer little control: a programmer cannot settle some decisions and have the autoscheduler complete the rest. 
Programmers thus choose between manual scheduling, which can reach any schedule but demands difficult reasoning about correctness and performance, and autoscheduling, which does not demand difficult human reasoning but often leaves performance on the table.\
More recently, large language models (LLMs) are used as aids to author schedules~\cite{lacouture:llm-autosparse, merouani:compilot, hong:llm-compilation, ramstad:llm-hls}, though their capabilities and limitations in this setting are less understood.



We propose scaffolding schedule authoring by offering guidance from multiple sources and guarantees of schedule validity. 
Experts understand many of the most important scheduling decisions well, such as which stages to fuse, while numerical choices like tile sizes are better delegated to automation. 
Autoschedulers also have different strengths; for example, one may vectorize effectively but miss opportunities to reuse data across stages. 
Ideally, schedule authors could \emph{compose} decisions from different autoschedulers and existing hand-tuned schedules, choosing the best source for each decision. 
Today, such composition is difficult. 
Autoschedulers can produce lengthy, complex schedules that are challenging to understand and impractical to modify. 
Moreover, because scheduling decisions are interdependent, validity constraints are often hard to reason about.

Our key insight is that compiler scheduling needs an abstraction that represents schedules as \emph{partial programs} to support both composing schedules from different sources and iteratively refining them into complete schedules. 
We take \emph{program sketches} as our abstraction and frame compiler scheduling as the iterative refinement of an \emph{underspecification} into a complete schedule. 
To make this refinement practical, we leverage Programming by Navigation (PBN)~\cite{lubin:honeybee}, an interactive synthesis technique. 
PBN guarantees that every choice it offers will lead to a valid schedule (\CStrongSoundness{}) and that all valid schedules are reachable (\CStrongCompleteness{}).

We present a methodology for interactively authoring compiler schedules by refining a program sketch whose holes represent scheduling decisions yet to be made.
We implement this method in \name{}, an interactive scheduling tool for two domain-specific compilers: Halide~\cite{ragan-kelley:halide} and FuseFlow~\cite{lacouture:fuseflow}.
At each round of interaction, the programmer fills a hole by selecting one of several steps offered by \name{}, each representing a single scheduling decision.
To guide these choices, \name{} \emph{ingests} existing schedules---whether generated by an autoscheduler or handwritten---and presents their decisions as \emph{hints} that the programmer can follow or ignore.
This design enables composition of scheduling decisions from autoschedulers, existing handwritten schedules, and the programmer.
In summary, our technical contributions are:
\begin{itemize}
    \item A method for interactively authoring compiler schedules with strong \emph{guarantees} by representing schedules as sketches and scheduling decisions as steps.
    \item \emph{Composition} of scheduling decisions drawn from autoschedulers, handwritten schedules, and the user by ingesting existing schedules and using them as sources of hints.
    \item Instantiations of \name{} for two domain-specific compilers, Halide~\cite{ragan-kelley:halide} and FuseFlow~\cite{lacouture:fuseflow}, demonstrating the generality of our approach across diverse domains and workloads.
\end{itemize}

We implement and evaluate \name{} for a subset of Halide's scheduling language and the full FuseFlow scheduling language. 
We show that schedules authored interactively with \name{} outperform the fastest existing schedules we evaluate, including autoscheduler-generated and handwritten schedules, by up to $1.8\times$ in Halide and $7.7\times$ in FuseFlow.
We further demonstrate that interactive synthesis scales to programs with hundreds of functions. 
Finally, LLM agents using \name{} reach high-performance FuseFlow schedules with 2.3$\times$ fewer output tokens on average than agents writing schedules directly.
More broadly, our results suggest that any compiler that exposes a scheduling language can support interactive schedule authoring that is valid by construction.

\section{Motivating Example}
\label{sec:motivating}

Many domain-specific compilers separate what a program computes from how it is computed. 
In Halide~\cite{ragan-kelley:halide}, a DSL for image processing, the \emph{algorithm} defines each stage of a pipeline as a pure function from pixel coordinates to values.
The \emph{schedule} defines how those values are computed by specifying the order of loops, which intermediate results are stored or recomputed, and which loops are vectorized or parallelized. 
\Cref{tab:halide-steps} summarizes the Halide scheduling operations used in this paper and their effects.
We use Halide as our primary domain-specific compiler example for this section, but the same separation appears in other scheduling languages~\cite{baghdadi:tiramisu, ikarashi:exo, ikarashi:exo2, koeplinger:spatial, lai:hetero-cl, chen:tvm, kjolstad:taco-sched, hsu:stardust, hsu:sam, hall:taxonomy, gyurgyik:scion, chen:allo, senanayake:sparse-iteration, hu:taichi, yadav:distal, yadav:spdistal, sundram:recuma, tariq:reptile, bansal:mosaic, hagedorn:rise-elevate, nayak:teaal, schulze:mdh, zhang:graphit, hagedorn:fireiron, fegade:cora, venkat:swirl, fegade:cortex}. 



Consider unsharp masking from Halide’s application suite~\cite{halide:repo}, which sharpens an image by converting it to grayscale, blurring it with a 7-tap Gaussian along each axis, and scaling each color channel by the ratio of the sharpened gray value to the original. 
The algorithm is:
\begin{lstlisting}[style=halide-algorithm,style=tight,label={fig:unsharp-alg}]
gray(x, y)      = 0.299f * input_bounded(x, y, 0) + 0.587f * input_bounded(x, y, 1)
                + 0.114f * input_bounded(x, y, 2);
blur_y(x, y)    = kernel(0) * gray(x, y) + kernel(1) * (gray(x, y - 1) + gray(x, y + 1)) + ...;
blur_x(x, y)    = kernel(0) * blur_y(x, y) + kernel(1) * (blur_y(x - 1, y) + blur_y(x + 1, y)) + ...;
sharpen(x, y)   = 2 * gray(x, y) - blur_x(x, y);
ratio(x, y)     = sharpen(x, y) / gray(x, y);
output(x, y, c) = ratio(x, y) * input(x, y, c);
\end{lstlisting}

Here, \code{kernel} holds the Gaussian weights, and \code{input_bounded} extends the image by repeating its edge pixels. 
This algorithm is short, but the schedule used to execute it drastically affects the runtime. 
On a 1536 $\times$ 2560 RGB image, the algorithm runs in 278.3\,ms with the default schedule. 
The expert handwritten schedule in the Halide repository~\cite{halide:repo}, shown in \Cref{fig:unsharp-schedules-hand}, runs in 1.02\,ms (a 273$\times$ speedup).
The schedule achieves this speedup by splitting the image into 32-row strips and processing the strips in parallel (lines 1--2).
Within each strip, it computes the intermediates \code{gray}, \code{blur_y}, and \code{ratio} one row at a time, right before each row is used (lines 3--5).
The schedule keeps the computed rows of each intermediate for the rest of the strip, so later output rows can reuse them instead of recomputing them. 

\begin{figure}[b]
\begin{centering}
\begin{minipage}{\linewidth}
\begin{lstlisting}[style=halide,style=tight]
output.split(y, y, yi, 20).split(x, x, xi, 512)
      .split(xi, xi, xii, 8).vectorize(xii)
      .reorder(xii, xi, (*\hlChan{yi, c}*), x, y).parallel(y);              (*\chg{1}*)
ratio.split(x, x, xi, 128)...(*\hlPlace{compute\_at(output, x)}*);       (*\chg{2}*)
gray...(*\hlPlace{compute\_at(ratio, x)}*);                              (*\chg{2}*)
blur_y...(*\hlPlace{compute\_at(ratio, xi)}*);                           (*\chg{2}*)
\end{lstlisting}
\subcaption{An excerpt of the 63-line schedule generated by the Adams2019 autoscheduler, which runs in $1.34$\,ms.}
\label{fig:unsharp-schedules-auto}
\end{minipage} \hfill
\begin{minipage}{\linewidth}
\begin{lstlisting}[style=halide,style=tight]
output.split(y, yo, yi, 32).vectorize(x, 8).parallel(yo)
      .reorder(x, (*\hlChan{c, yi}*), yo);                                  (*\chg{1}*)
gray.(*\hlPlace{compute\_at(output, yi).store\_at(output, yo)}*).vectorize(x, 8);    (*\chg{2}*)
blur_y.(*\hlPlace{compute\_at(output, yi).store\_at(output, yo)}*).vectorize(x, 8);  (*\chg{2}*)
ratio.(*\hlPlace{compute\_at(output, yi).store\_at(output, yo)}*).vectorize(x, 8);   (*\chg{2}*)
\end{lstlisting}
\subcaption{The handwritten schedule from the Halide repository, which runs in $1.02$\,ms.} 
\label{fig:unsharp-schedules-hand}
\end{minipage}
\end{centering}
\caption{
Two Halide schedules, authored differently, for unsharp masking (algorithm above in \Cref{fig:unsharp-alg}).
We highlight the two main differences between their scheduling decisions:
\chg{1}~places the channel loop \code{c} outside or inside the row loop \code{yi};
\chg{2}~sets how much of each intermediate is computed and stored at once.
\Cref{fig:unsharp-schedules-auto} computes \code{ratio} per $512 \times 20$ block, \code{gray} per 128-column chunk of \code{ratio}, and \code{blur\_y} per 64-column piece of a row, whereas \Cref{fig:unsharp-schedules-hand} computes all three one row at a time and stores them for the whole strip.
}
\label{fig:unsharp-schedules}
\end{figure}

The Halide codebase ships with three autoschedulers for CPUs: Li2018~\cite{li:differentiable-halide}, Mullapudi2016~\cite{mullapudi:halide-auto}, and Adams2019~\cite{adams:learning}. 
Their schedules for unsharp masking run in 9.87, 1.84, and 1.34\,ms, respectively.
We focus on Adams2019, the fastest, and show part of its schedule in \Cref{fig:unsharp-schedules-auto}.
Like the handwritten schedule, the Adams2019 schedule runs many small tasks in parallel; however, it differs in two decisions that account for the remaining gap in their runtimes.
First, the \code{reorder} on line~3 of the Adams2019 schedule in \Cref{fig:unsharp-schedules} places the channel loop outside the row loop, so Adams2019 cannot efficiently compute an intermediate one row at a time.
Second, the \code{compute_at} directives on lines~4--6 compute the intermediates in overlapping chunks, so Adams2019 recomputes the rows and columns that neighboring chunks share.

Although a few coordinated changes make the Adams2019 schedule as fast as the handwritten schedule, identifying and applying those changes in the generated code is difficult. 
In order to alter the Adams2019 schedule to achieve the handwritten schedule's performance, we must move the channel loop inside the row loop and compute intermediates one row at a time. 
These changes interact with decisions elsewhere in Adams2019's schedule.
For example, removing \code{ratio}'s \code{xi} chunk loop leaves \code{blur\_y.compute\_at(ratio, xi)} referring to a loop that no longer exists, so its placement must also change. 
Further, while moving the channel loop alone is valid, it does not improve runtime until the intermediate computations also change. 
Due to these complexities, experts in practice typically author a schedule fully manually rather than attempt to modify an autoscheduler's output. 
How, then, can the programmer build on Adams2019's useful decisions without manually untangling the decisions in the generated code?



\section{System Overview}
\label{sec:overview}
\begin{figure}[t]
  \centering
  \includegraphics[width=\linewidth]{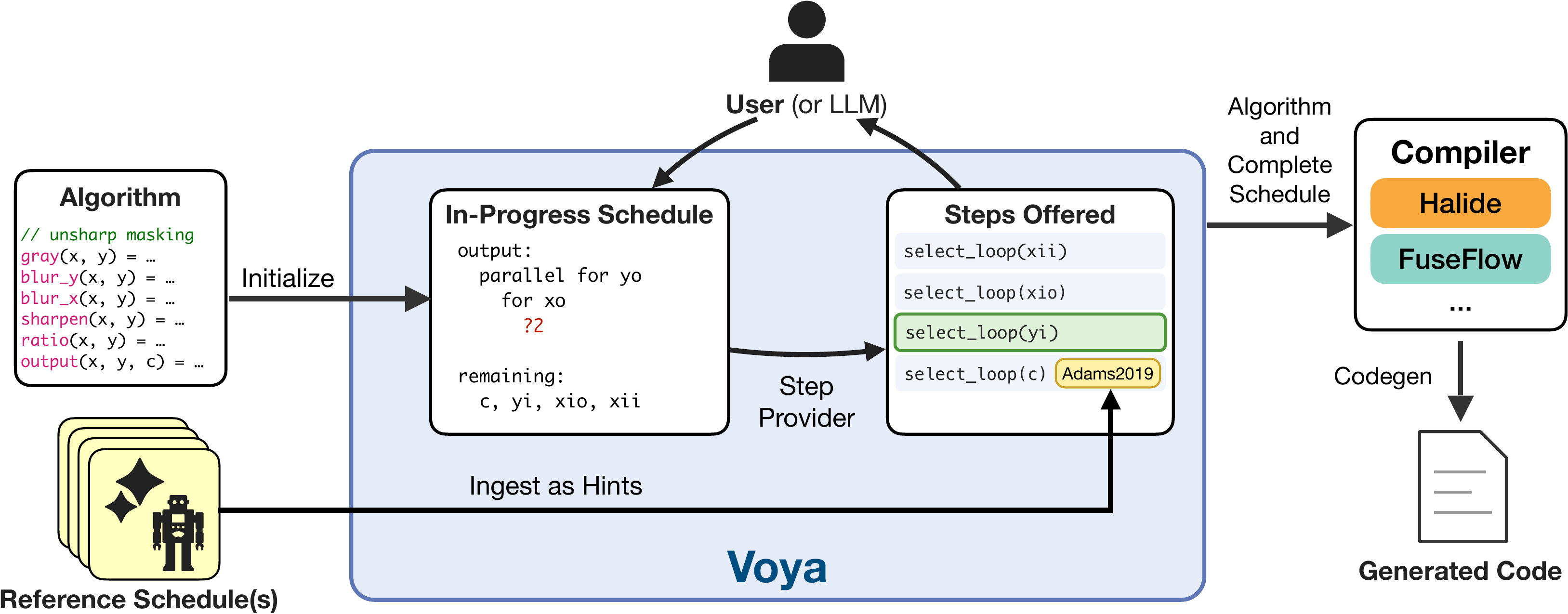}
  \caption{
  Overview of \name{}.
  \name{} parses the input algorithm to initialize a working sketch with holes.
  At each round of interaction, the step provider offers a set of steps to the user.
  The user (a programmer or LLM agent) selects one, which fills a hole and may introduce new ones.
  Reference schedules, either handwritten or autoscheduler-generated, can optionally be ingested by \name{} to provide hints during the interaction. 
  When all holes are filled, the complete schedule and input algorithm are passed to the compiler.
  }
  \label{fig:hbsched-overview}
\end{figure}
\name{} is an interactive tool for authoring compiler schedules built using Honeybee~\cite{lubin:honeybee}, a Programming by Navigation synthesizer. 
\name{} takes in an \emph{algorithm} written in a DSL, and launches an interactive session with a user to construct a schedule for that algorithm.
\Cref{fig:hbsched-overview} shows this framework for our two compiler instantiations, Halide and FuseFlow. 
Each interactive session launched can be driven by either a human or LLM user, or both. 

Information extracted from the input algorithm on the left of \Cref{fig:hbsched-overview}, such as its stages and their dependencies, is used to initialize the schedule sketch with holes inside \name{} as an initial in-progress schedule (working sketch).
The compiler-specific scheduling library defines the steps that fill these holes, and applying a step may introduce further holes.  
The arrows between \name{} and the user show the CLI presenting the working sketch and available next steps for the user to select.
\name{} repeats this interaction until the schedule is complete, then passes it with the algorithm to the compiler on the right.

Existing schedules, whether handwritten or generated by autoschedulers, can optionally be ingested by \name{} to provide hints during the interactive process.
Corresponding offered steps are marked with their source schedules, allowing the user to follow or ignore their recommendations.
Autoschedulers run on the same input algorithm, following the lower path in \Cref{fig:hbsched-overview}.
\Cref{sec:pbn} describes this scheduling representation and the theory behind schedule ingestion. 

Our implementation extends Honeybee's synthesis engine, with libraries defining compiler-specific scheduling operations, constraints, and code generation.
See \Cref{tab:halide-steps} and \Cref{tab:fuseflow-steps} for the scheduling operators we support in Halide and FuseFlow, respectively.  
\Cref{sec:instantiations} details how we implement the instantiations to ensure forward progress, enforce phase-ordering constraints, and handle domain-specific differences.   

\subsection{System Walkthrough}
We return to the unsharp masking kernel of \Cref{sec:motivating} to demonstrate how a programmer interacts with \name{} to construct a complete schedule. 
The programmer’s goal is to retain Adams2019’s blocking and parallelization while computing intermediates one row at a time and reusing them across neighboring rows.
\name{} takes the unsharp masking algorithm and initializes an in-progress schedule, or \textit{sketch}, with holes standing in for the scheduling decisions the programmer has yet to make. 
Supplying Adams2019's schedule as a source of \textit{hints} allows the programmer to reuse its decisions while choosing alternatives where they want to intervene. 

\begin{figure}
\begin{centering}
\begin{minipage}{0.43\linewidth}
\begin{lstlisting}[style=menu,escapechar=!]
Working schedule:
build_pipeline(
  output = compute_root(?1), // nest: x y c
  ...)

Possible next steps (excerpt):
?1 !$\mapsto$! split(y, 2)
?1 !$\mapsto$! split(y, 4)
...
?1 !$\mapsto$! split(y, 20)         !$\leftarrow$! Adams2019
...
\end{lstlisting}
\subcaption{Round 1}
\label{fig:interaction-r1}
\end{minipage} \hfill
\includegraphics[scale=0.3,angle=90]{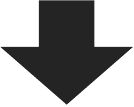} \hfill
\begin{minipage} {0.5\linewidth}
\begin{lstlisting}[style=menu,escapechar=!]
Working schedule:
output: yo (parallel) -> xo -> ?2
Remaining loops: c, yi, xio, xii

Possible next steps (excerpt):
?2 !$\mapsto$! select_loop(c)        !$\leftarrow$! Adams2019
?2 !$\mapsto$! select_loop(yi)       // programmer selects
?2 !$\mapsto$! select_loop(xio)
?2 !$\mapsto$! select_loop(xii)
...
\end{lstlisting}
\subcaption{Round 2}
\label{fig:interaction-r2}
\end{minipage}
\end{centering}
\caption{Two rounds of interactive scheduling with \name{} for Halide's unsharp masking.}
\label{fig:interaction}
\end{figure}

These hints appear alongside the steps that \name{} offers at each
round of interaction. 
For example, after choosing to compute \code{output} at the root, the programmer encounters the following choices, shown in \Cref{fig:interaction-r1} for splitting its loops. 
Following Adams2019's hint, the programmer chooses \code{split(y, 20)}, dividing \code{y} into an outer loop \code{yo} over blocks and an inner loop \code{yi} over rows within each block.
Further hints guide the split of \code{x} into 512-column blocks and the subdivision of each block into groups of eight columns for vectorization.
Placing \code{yo} outermost, parallelizing it, and placing \code{xo} immediately inside that loop recreates the blocking and parallelization strategy of Adams2019.
The next decision, shown in \Cref{fig:interaction-r2}, is where to depart from its recommendations.

Here, Adams2019 recommends selecting the loop over \code{c}, but the desired row-by-row computation requires \code{yi} to come first.
Taking this unmarked alternative nests \code{c} inside the loop of \code{yi}.
Choosing \code{compute_at(output, yi)} and \code{store_at(output, xo)} for \code{ratio}, \code{gray}, and \code{blur_y} in subsequent menus completes the intended changes.
\name{} enables the programmer to compose Adams2019's blocking and parallelization decisions with their own choices for loop order and intermediate placement to obtain a faster schedule.



\section{Compiler Scheduling as Interactive Specification Refinement}
\label{sec:pbn}

To precisely state the guarantees \name{} offers, we first formalize our problem statement. We start with some preliminary definitions to model compiler scheduling languages. 
A schedule fixes decisions such as the iteration order, the storage granularity, and the degree of parallelism.
Given an algorithm and a schedule, a compiler either produces a target program or rejects the schedule (which we model as returning $\bot$), \textit{e.g.} when the schedule refers to a loop that does not exist. We capture this behavior independently of any particular compiler with the following formal definition.

\begin{definition}[Schedules]
\label{def:schedules}
For a given set of \textbf{source programs} $ \Src $, \textbf{target programs} $ \Tgt $, we will consider \textbf{schedules} $ \Exp $ whose semantics are partial functions $\Denot{\Exp} : \Src \rightharpoonup \Tgt$. We call a program $ \Alg $ in $ \Src $ an \textbf{algorithm} and say that a schedule $\Exp $ is \textbf{valid} for $\Alg $, written $\ValidA{\Exp}$, if $\Denot{\Exp}\,\Alg \neq \bot$.
\end{definition}

The \emph{compiler scheduling problem} is to produce a schedule $ \Exp $ with $\ValidA{\Exp}$. 
However, as we discuss in Section~\ref{sec:intro}, this requirement is a vast underspecification, as for realistic algorithms there can be millions of different possible schedules; which one is best depends on the task at hand and, often, the input data.
%
%
Our goal, then, is to provide a way for users to refine their initial underspecification interactively. 
To do so, we will take our notion of schedules to be \emph{program sketches}~\cite{solar-lezama:sketching}, partially-constructed programs in which remaining decisions are \textit{holes}, written $ \Hole{\HName} $.


\begin{definition}[Sketches]
\label{def:sketches}
We take our schedules to be \textbf{sketches} $ \Exp $, expressions composed of $N$-ary scheduling operations $ \Fun $ with holes $ \Hole{\HName} $ defined by $
    \Exp \Sis \App{\Fun}{\ExpList} \Sor \Hole{\HName}
$ where $ N \geq 0 $.
We require that the subscript $\HName$ uniquely names each hole in an expression.
We say a sketch with no holes is \textbf{complete}, and the \textbf{blank schedule} is a single hole, $\Start = \Hole{0}$.
\end{definition}

We will then allow these sketches to be filled out using top-down refinement steps that either replace a hole with a function application or are a sequence of such steps. These steps will be provided to the user to select between to refine their schedule.


\begin{definition}[Top-down refinement step]
\label{def:topdown}
A \textbf{top-down refinement step} $\Step$ is defined by $\Step \Sis \StepTD{\HName}{\App{\Fun}{\HoleList{\HName}}} \Sor \StepSeq{\StepSub{1}}{\StepSub{2}} $ with $N \geq 0$. We
say $\ExpSub{1}$ \textbf{steps to} $\ExpSub{2}$ (written $\StepD{\ExpSub{1}}{\Step}{\ExpSub{2}} $) for $ \Step = \StepTD{\HName}{\App{\Fun}{\HoleList{\HName}}} $ if $\Hole{\HName}$ is a subexpression of $ \ExpSub{1} $ and $\ExpSub{2}$ is
obtained by replacing $\Hole{\HName}$ with
$\App{\Fun}{\HoleList{\HName}}$ in  $\ExpSub{1}$, where $\HoleList{\HName}$ are pairwise distinct and not previously used in the interaction; for $ \Step = \StepSeq{\StepSub{1}}{\StepSub{2}} $, we say $\StepD{\ExpSub{1}}{\Step}{\ExpSub{2}}$ if there exists an expression $ \Exp{} $ such that $\StepD{\ExpSub{1}}{\StepSub{1}}{\Exp} $ and $\StepD{\Exp}{\StepSub{2}}{\ExpSub{2}} $.
\end{definition}

Interactive refinement, however, introduces a new problem. 
Validity is defined only on complete schedules, so during refinement, the compiler has no feedback to provide about a partially constructed sketch. 
Further, a user may go down a ``rabbit hole'' of steps that lead to no valid schedule. 
Withholding steps conservatively is also problematic, as an overcautious tool may render the schedule the user wants unreachable.

To avoid both (i)~showing steps that do not lead to a valid solution and (ii)~not showing enough steps to lead to all valid solutions, we cast the compiler scheduling problem as an instance of Programming by Navigation (PBN) \cite{lubin:honeybee}, an interactive specification refinement technique. Given our notion of expressions and steps above, Programming by Navigation specification refinement works by having a \textit{step provider} offer a set of steps $ \StepSet $ that a \textit{step decider} (\textit{e.g.}, a human or automated procedure) can select between. Crucially, in PBN, this step set $ \StepSet $ must satisfy \CStrongSoundness{} (all provided steps must lead to a valid schedule) and \CStrongCompleteness{} (any remaining valid schedule is reachable among the provided steps), precisely avoiding the two issues above. We will implement such a step provider, solving our formulation of the compiler scheduling problem for two different scheduling languages, in Section~\ref{sec:instantiations}.

\begin{figure}[t]
\centering
\resizebox{\linewidth}{!}{%
\begin{forest}
  sketch/.style={
    draw=pathgreen, fill=pathgreen!10,
    rounded corners, font=\ttfamily\scriptsize,
    align=center, inner sep=4pt
  },
  step/.style={
    draw=solarized-blue, rounded corners,
    font=\ttfamily\scriptsize, align=center, inner sep=4pt
  },
  chosen/.style={
    step, draw=pathgreen, fill=pathgreen!10,
    line width=1pt,    
    edge={->,>=stealth,pathgreen,line width=1pt}
  },
  pruned/.style={
    draw=solarized-base1, text=solarized-base1,
    rounded corners, font=\ttfamily\scriptsize,
    align=center, inner sep=4pt,
    edge={->,>=stealth,solarized-red,densely dashed},
    edge label={
      node[midway,right,solarized-red]{$\times$}
    }
  },
  dots/.style={
    draw=none, text=solarized-base1, font=\normalsize
  },
  for tree={
    edge={->,>=stealth,solarized-blue},
    l sep=5mm, s sep=3mm
  },
  [{output.$\Hole{1}$}, sketch
    [{$\sigma_1:\ \Hole{1}\rightsquigarrow$\\
      split(y, yo, yi, 32).$\Hole{2}$}, chosen
      [{$\sigma_2:\ \Hole{2}\rightsquigarrow$\\
        vectorize(x, 8).$\Hole{3}$}, chosen
        [{$\sigma_3:\ \Hole{3}\rightsquigarrow$\\
          parallel(yo)}, chosen, name=s3]
        [{$\Hole{3}\rightsquigarrow$\\
          serial(yo)}, step]
        [{$\cdots$}, dots]
      ]
      [{$\Hole{2}\rightsquigarrow$\\
        serial(x).$\Hole{3}$}, step
        [{$\vdots$}, dots]
      ]
      [{$\cdots$}, dots]
    ]
    [{$\Hole{1}\rightsquigarrow$\\
      split(y, yo, yi, 20).$\Hole{2}$}, step
      [{$\vdots$}, dots]
    ]
    [{$\Hole{1}\rightsquigarrow$\\
      parallel(yo).$\Hole{2}$}, pruned]
    [{$\cdots$}, dots, name=rdots]
  ]
  \node[anchor=south east, font=\ttfamily\scriptsize, inner sep=0pt,
        yshift=-0.5ex]
    at (s3.south -| rdots.east) {%
    \renewcommand{\arraystretch}{1.15}%
    \begin{tabular}{@{}r@{\ }l@{}}
      & \lstinline[style=halide, basicstyle=\ttfamily\scriptsize]{output.}$\Hole{1}$\\
      $\xrightarrow{\StepSub{1}}$
      & \lstinline[style=halide, basicstyle=\ttfamily\scriptsize]{output.split(y,yo,yi,32).}$\Hole{2}$\\
      $\xrightarrow{\StepSub{2}}$
      & \lstinline[style=halide, basicstyle=\ttfamily\scriptsize]{output.split(y,yo,yi,32).vectorize(x,8).}$\Hole{3}$\\
      $\xrightarrow{\StepSub{3}}$
      & \lstinline[style=halide, basicstyle=\ttfamily\scriptsize]{output.split(y,yo,yi,32).vectorize(x,8).parallel(yo)}
    \end{tabular}};
\end{forest}}
\caption{
Refining a fragment of unsharp masking's handwritten schedule.
The root is the current sketch; each subsequent node is a step
that fills a hole and may introduce a fresh hole.
The highlighted path selects three steps, producing the sketches on the bottom right.
Other branches show alternative choices.
The dashed step is withheld because \texttt{yo} does not exist
before the split.
Every offered step admits a valid completion
(\CStrongSoundness{}), and every remaining valid schedule is
reachable through offered steps (\CStrongCompleteness{}).
Ellipses indicate additional alternatives or further steps;
\texttt{serial} selects a loop's default execution. 
Steps shown use Halide's chained syntax. 
$\StepTD{1}{\texttt{split(y, yo, yi, 32)}.\Hole{2}}$ is the step $\StepTD{1}{\App{\Fun}{\Hole{2}}}$ of Definition~\ref{def:topdown} with $\Fun = \texttt{split(y, yo, yi, 32)}$, whose arguments are parameters rather than holes.
}
\label{fig:space}
\label{fig:unsharp-refinements}
\end{figure}
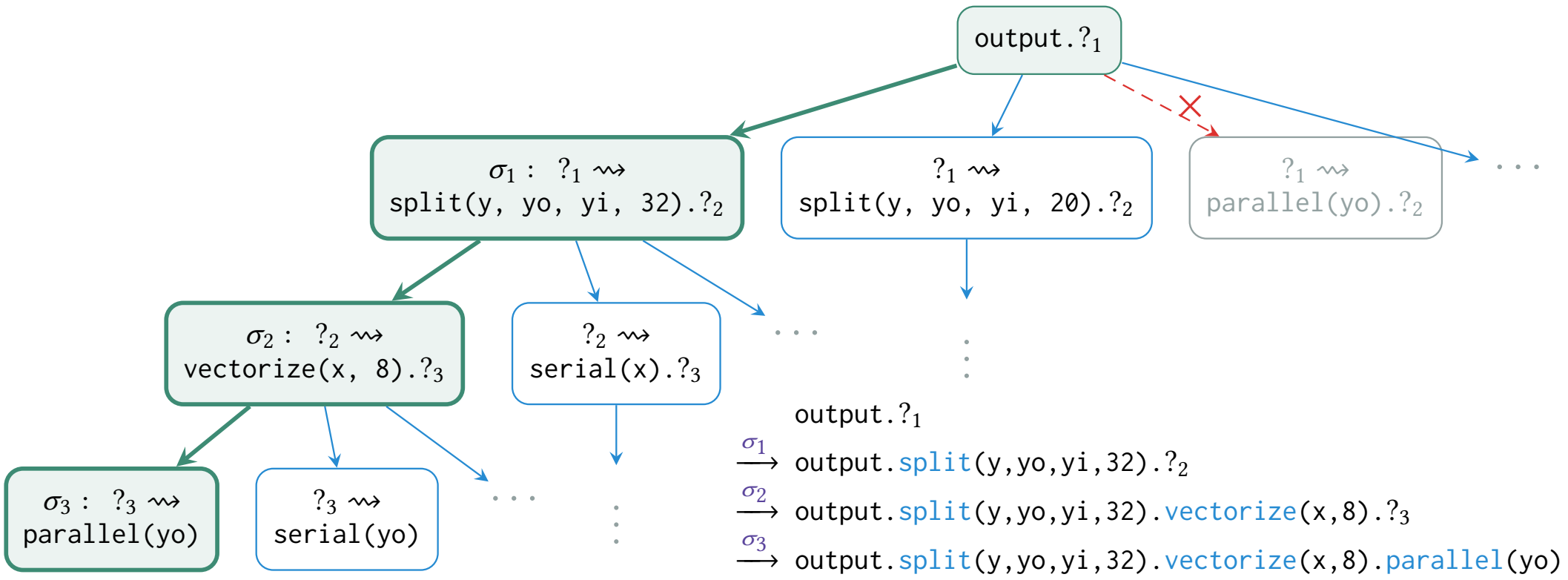

As an example, consider again unsharp masking from \Cref{sec:motivating}.
\Cref{fig:space} shows a fragment of the choices for scheduling the function \code{output}, focusing on its loop nest. Starting from \code{output} (visualized at the top), a step provider provides a set of next steps $ \StepSetSub{0} $ that contains steps like $ \StepSub{1} $ (which are en route to a valid solution), but not any of the steps that cannot lead to a valid solution. The highlighted path follows three steps from the handwritten schedule in \Cref{fig:unsharp-schedules}: first splitting \code{y} into 32-row strips~($\StepSub{1}$), then vectorizing \code{x} by eight~($\StepSub{2}$), and finally parallelizing the outer loop \code{yo}~($\StepSub{3}$).
The refinements below the tree show how each selected step fills a hole and introduces fresh holes as needed for the remaining decisions. Our goal will be to define the step provider that provides this set of steps for a step decider to select between.

\subsection{Hints}
\label{sec:hints}
In the previous section, we saw how Programming by Navigation step providers offer guarantees on which refinement steps are provided to the step decider. However, they do not \emph{guide} the user in deciding which steps to take. 
To guide the step decider's choice among the offered steps, we offer a notion of \emph{hints}.
Intuitively, hints should be able to come from an autoscheduler, a human expert, or a cost model;
a source may recommend some steps strongly and others weakly, and different sources may disagree.
Formally, we express a hint as a probability distribution over the steps provided by a Programming by Navigation step provider at each hole.

\begin{definition}[Hints]
\label{def:hints}
Let $\StepSet$ be a set of steps provided by a step provider. 
A \textbf{hint} for $\StepSet$ is a
probability distribution over $\StepSet$, that is, a function
$ \Hint : \StepSet{} \to [0, 1] $ with 
$ \sum_{\Step \in \StepSet} \Hint(\Step) = 1.
$
\end{definition}

A hint weights the offered steps but never adds or removes one, so hints preserve \CStrongSoundness{} and \CStrongCompleteness{}. The step decider may follow a hint or ignore it.

\begin{remark}
Hints can be richer than a single recommendation, since we may weight the steps at a hole by a cost model's estimates or by a language model's confidence in each alternative.
Therefore, it can be useful to compose hints from different sources.
Given hints $\Hint_1, \dots, \Hint_m$ for the same hole, we combine them by aggregating their distributions pointwise, for example, by averaging and renormalizing.
The result is again a hint. 
When multiple sources recommend the same step, the weights accumulate on the recommended step. 
When they disagree, the step decider sees the alternatives weighted side by side and can choose among them.
\end{remark}


As we explore empirically in \Cref{sec:eval-rq3}, the above framing of scheduling using Programming by Navigation along with hints suggests a new approach to \textit{autoscheduling}, in which an autoscheduler is a procedure that need only \textit{select between} (possibly-hinted) steps provided by a step provider. \CStrongSoundness{} guarantees that any program produced by such an autoscheduler will be valid, and \CStrongCompleteness{} guarantees that all programs are accessible to the autoscheduler.


\subsection{Schedules as Traces}
\label{sec:ingest}

For now, though, the last question we must answer is: Where should we obtain hints \textit{from}? It turns out that any method that produces a schedule (including manual, handwritten schedules and schedules produced by an autoscheduler) can be transformed into a source of hints on steps.

To derive hints from a reference schedule, we must recover the steps taken to construct it.
Consider a simple schedule for unsharp masking from
\Cref{sec:motivating}, showing only \code{output} and \code{ratio}:
\[
  \Exp = \mathtt{build\_pipeline}(
    \mathtt{compute\_root}(\mathtt{serial}),\;
    \mathtt{compute\_at}(\mathtt{serial})).
\]
Here, \code{build_pipeline} groups the schedules for \code{output} and \code{ratio}, respectively.
This schedule computes \code{output} at the root and \code{ratio} inside \code{output}'s \code{y} loop.
The terminal operation \code{serial} closes a hole without adding any loop scheduling directives, leaving Halide's default loop behavior unchanged.
(For brevity, we omit the schedules for the arguments and other functions in the algorithm.)
Given our notion of steps, every interaction that reaches $\Exp$ begins with $\StepTD{0}{\mathtt{build\_pipeline}(\Hole{1}, \ldots, \Hole{n})}$, introducing one hole for each function's schedule:

\begin{center}
\begin{tikzpicture}[
    every node/.style={font=\footnotesize},
    lbl/.style={font=\scriptsize, midway}]
  \node (top) at (0, 0)
    {$\mathtt{build\_pipeline}(\Hole{1}, \Hole{2})$};
  \node[align=center] (l) at (-2.4, -1.6)
    {$\mathtt{build\_pipeline}($\\
     $\mathtt{compute\_root}(\Hole{3}), \Hole{2})$};
  \node[align=center] (r) at (2.4, -1.6)
    {$\mathtt{build\_pipeline}($\\
     $\Hole{1}, \mathtt{compute\_at}(\Hole{3}))$};
  \node (bot) at (0, -3.2) {$\Exp$};

  \draw[->] (top) -- node[lbl, above left]
    {$\StepTD{1}{\mathtt{compute\_root}(\Hole{3})}$} (l);
  \draw[->] (top) -- node[lbl, above right]
    {$\StepTD{2}{\mathtt{compute\_at}(\Hole{3})}$} (r);
  \draw[->] (l) -- node[lbl, below left, align=center]
    {$\StepSeq{\StepTD{2}{\mathtt{compute\_at}(\Hole{4})}}{}$\\
     $\StepSeq{\StepTD{3}{\mathtt{serial}}}
              {\StepTD{4}{\mathtt{serial}}}$} (bot);
  \draw[->] (r) -- node[lbl, below right, align=center]
    {$\StepSeq{\StepTD{1}{\mathtt{compute\_root}(\Hole{4})}}{}$\\
     $\StepSeq{\StepTD{3}{\mathtt{serial}}}
              {\StepTD{4}{\mathtt{serial}}}$} (bot);
\end{tikzpicture}
\end{center}

Both interactions reach $\Exp$, and their steps differ only because the user filled the holes in a different order.
However, we can see that, up to exchanging $\Hole{3}$ and $\Hole{4}$, the two interactions take the same \textit{set} of steps, which motivates the following definition.

\begin{definition}[Traces]
The \textbf{trace} of an interaction
$
 \StepD{\Start}{\StepSub{1}}{\StepD{\cdots}{\StepSub{N}}{\Exp}}
$ where each $ \StepSub{i} $ is a hole-filling step,
is the unordered set $\TraceSet = \{\StepSub{1}, \ldots, \StepSub{N}\}$.
A trace of a complete schedule $\Exp$ is the trace of any such interaction
from $\Start$ to $\Exp$. 
Two traces are \textbf{equivalent up to hole renaming} when there exists a bijection of hole names that maps one onto the other.
\end{definition}

Given a trace, we can define a hint for how to fill a hole. At a hole $\Hole{\HName}$ of the working sketch, if the trace fills the corresponding hole with an operation $\Fun$, the corresponding hint is to place probability $1$ on the offered step that fills $\Hole{\HName}$ with $\Fun$.
If no such step is offered (\textit{e.g.} because the user has departed from the reference earlier in the interaction), the trace simply offers no hint. As the earlier example suggests, schedules uniquely determine the hints produced by traces in such a manner, which is a consequence of the following theorem.

\begin{theorem}[Trace uniqueness]
\label{thm:uniqueness}
All traces of a complete schedule $ \Exp $ are equivalent up to hole renaming.\footnote{Not all instantiations of Programming by Navigation satisfy analogous properties to \textsc{Trace Uniqueness}, but our particular choice of top-down refinement steps does.
Consider expressions $\mathbb{N}$ with steps of the form $+k$ ($k \geq 1$): $\StepD{0}{+3}{3}$ and $\StepD{0}{+1}{\StepD{1}{+2}{3}}$
reach the same endpoint with distinct traces $\{+3\}$ and $\{+1, +2\}$.}
\end{theorem}

\begin{proof}
Each top-down step substitutes at a single hole, so it introduces exactly one operation occurrence of the final schedule and leaves every occurrence already present unchanged.
Since hole names are never reused, distinct steps of an interaction fill distinct holes.
Mapping each step of an interaction to the occurrence it introduces is therefore a bijection from its trace onto the operation occurrences of $ \Exp $.
Hence the traces of any two interactions reaching $ \Exp $ correspond to the same occurrences, and are thus equivalent up to the renaming of holes.
\end{proof}



Finally, we can define a syntax-directed procedure that \emph{ingests} an existing schedule to produce a trace (for use in hinting).
Given a schedule $ \Exp $ and the name $ \HName $ of the hole it fills, it returns the set of steps $\TraceSet$ that construct $\Exp$ from $ \Hole{\HName} $. We write this as $\Ingest{\HName}{\Exp}{\TraceSet}$:

\begin{mathpar}
\inferrule*[right=\textsc{Ingest/Hole}]
  { }
  {\Ingest{\HName}{\Hole{\HName'}}{\varnothing}}

\inferrule*[right=\textsc{Ingest/App}]
  {\forall i.\; \Ingest{\HName_i}{\ExpSub{i}}{\TraceSetSub{i}} \\
   \HName_1, \ldots, \HName_N \text{ fresh}}
  {\Ingest{\HName}{\App{\Fun}{\ExpList}}
          {\{\StepTD{\HName}{\App{\Fun}{\HoleList{\HName}}}\}
           \cup \textstyle\bigcup_i \TraceSetSub{i}}}
\end{mathpar}

Exactly one rule applies to each form of $\Exp$, and each premise is on
a strict subterm, so ingestion runs in a single pass over $\Exp$.
Fresh hole names are distinct from each other and from $ \HName $ across a derivation, so no two steps fill the same hole. 
\textsc{Ingest/Hole} leaves a hole unfilled, which lets us ingest
partial schedules, such as those from autoschedulers that decide only
part of a schedule.
The trace of a complete schedule $\Exp$ is the trace 
$\TraceSet$ with $\Ingest{0}{\Exp}{\TraceSet}$, starting from the blank
sketch $\Start = \Hole{0}$.

\begin{theorem}[Ingestion]
\label{thm:ingest}
Let $ \Exp $ be a schedule with $\Ingest{\HName}{\Exp}{\TraceSet}$. 
Then $ \TraceSet $ is the trace of an interaction from $ \Hole{\HName}$ to a sketch equal to $ \Exp $ up to hole renaming.
\end{theorem}

\begin{proof}
By induction on $\Exp$.
\end{proof}

Combining Theorems~\ref{thm:uniqueness} and~\ref{thm:ingest}, every complete schedule has a trace that is unique up to hole renaming and computed in a single pass, however the schedule was produced. 
Therefore, a reference schedule, such as an autoscheduler's output or an expert's handwritten schedule, yields a hint (Definition~\ref{def:hints}) through its trace.

\section{Instantiating \name{} in Halide and FuseFlow}
\label{sec:instantiations}

A \name{} instantiation for a particular scheduling language requires (i)~a library of scheduling operations expressed as top-down steps that fill holes in the working sketch, and (ii)~a \textit{step provider} that, given a working sketch, returns the set of steps from which a valid schedule is still reachable.
Given the library and step provider, \CStrongSoundness{} and \CStrongCompleteness{} follow directly and no additional language-specific work is required to support schedule ingestion or hints. 
We emphasize that the development of \Cref{sec:pbn} applies to any scheduling language whose operations can be expressed as top-down steps (Definition~\ref{def:topdown}).

Implementing a step provider for a scheduling language poses two key challenges. 
First, scheduling transformations can be cyclic, so the step provider must withhold steps that revisit previously seen states to ensure forward progress.
Second, in the presence of scheduling choices that are valid locally but invalid globally, the step provider must decide whether a valid schedule is still reachable. 
We discuss forward progress in \Cref{sec:instantiations-structure} and constraint validation in \Cref{sec:instantiations-validity}. 
We address both challenges in two domain-specific compilers with very different scheduling languages. 
Halide~\cite{ragan-kelley:halide} compiles dense image processing pipelines for CPUs and GPUs, with schedules that transform each function's loop nest.
FuseFlow~\cite{lacouture:fuseflow} compiles sparse deep-learning models to streaming dataflow accelerators~\cite{hsu:sam, zhang:dam, koul:onyx,koul:onyx-jssc}, and its schedules choose each
tensor's sparse storage format and which operations fuse into a single dataflow kernel.
We support the Halide subset in \Cref{tab:halide-steps} and the full FuseFlow scheduling language (\Cref{tab:fuseflow-steps}). 
\CStrongCompleteness{} holds with respect to these scheduling spaces.




\subsection{Representing Scheduling Operations as Steps}
\label{sec:instantiations-steps}

To add a scheduling language to \name{}, we implement a library that encodes its scheduling operations as steps.
When scheduling decisions are independent, the user may make them in any order.
Some decisions, however, depend on others.
Generally in loop-based languages, \code{split} and \code{fuse} directives can create loops that later \code{unroll} or \code{vectorize} directives refer to.
The library enforces these orderings, for example by requiring that steps that create loops be chosen before steps that refer to them.
We now describe the libraries for Halide and FuseFlow.

\paragraph{Halide}
A Halide pipeline is a directed acyclic graph of functions, where each function's values are defined in terms of the functions it consumes. 
In unsharp masking (\Cref{sec:motivating}) for example, \code{gray} is consumed by \code{blur_y}, \code{sharpen}, and \code{ratio}.
Every interaction begins with the step $\StepTD{0}{\mathtt{build\_pipeline}(\Hole{1}, \ldots, \Hole{n})}$, which introduces one hole per function. 
\Cref{tab:halide-steps} lists the scheduling operations we support. 
Those above the line decide where each function is computed and stored, while those below the line transform each function's loop nest.
A step fills a hole with one operation and introduces a fresh hole for the function's remaining decisions.
\Cref{fig:space} shows how repeated steps grow the schedule of \code{output} into a chain of operations.

\begin{table}[t]
  \centering
  \small
  \caption{The Halide scheduling operations represented as \name{} steps.
  Operations below the line apply to a function's pure definition and to each of its update definitions.
  }
  \label{tab:halide-steps}
  \begin{tabularx}{\linewidth}{@{}l X@{}}
    \toprule
    \textbf{Scheduling Operation} & \textbf{Effect} \\
    \midrule
    \texttt{compute\_root}    & compute the function in full before its consumers \\
    \texttt{compute\_at}      & compute the function inside a consumer's loop \\
    \texttt{store\_at}        & store a computed-at function at an outer loop (sliding window) \\
    \texttt{compute\_inline}  & recompute the function at each use, without separate storage \\
    \texttt{store\_in}        & place the function's buffer on the stack instead of the heap \\
    \texttt{reorder\_storage} & make a chosen dimension of the buffer innermost \\
    \midrule
    \texttt{split}            & divide a loop into an outer and an inner loop \\
    \texttt{fuse}             & combine two adjacent loops into one \\
    \texttt{reorder}          & choose the order of a nest's loops \\
    \texttt{vectorize}        & map a loop to SIMD lanes \\
    \texttt{parallel}         & run a loop across threads \\
    \texttt{unroll}           & unroll a loop \\
    \texttt{serial}           & run a loop sequentially (default behavior) \\
    \bottomrule
  \end{tabularx}
\end{table}




\paragraph{FuseFlow}
FuseFlow compiles sparse deep-learning models to sequences of sparse tensor algebra expressions in Einsum notation~\cite{kjolstad:taco}.
These expressions are lowered to fused streaming dataflow graphs.  
As a running example, we use the chained matrix product $D_{il} = A_{ij}B_{jk}C_{kl}$, which multiplies three (potentially sparse) matrices into the output $D$.
\Cref{tab:fuseflow-steps} lists FuseFlow's scheduling operations, which choose a format for each tensor, decide at each boundary between operations whether to fuse the two sides into one kernel, and set each kernel's loop order, parallelization, stream shape, and blocking.
\name{} initializes the sketch with a hole for each tensor's format and each boundary between operations. 
We refer readers to \citet{lacouture:fuseflow} for further details on these scheduling operations.
Once the fusion plan is fixed, each resulting kernel receives holes for its loop order, parallelization, stream shape, and blocking.
For our example, the sketch begins with format holes for $A$, $B$, $C$, and $D$. 
\Cref{fig:ff-anno} shows the steps offered for $A$ once the other three are filled.

\begin{table}[tbp]
  \centering
  \small
  \caption{
    The FuseFlow scheduling operations represented as \name{} steps.
    Each tensor receives one storage format.
    Each boundary between operations receives either \texttt{cut\_after\_op} or \texttt{extend\_across}, deciding whether the operations on either side are fused.
    The remaining steps apply to each fused kernel. 
  }
  \label{tab:fuseflow-steps}
  \begin{tabularx}{\linewidth}{@{}l X@{}}
    \toprule
    \textbf{Scheduling Operation} & \textbf{Effect} \\
    \midrule
    \texttt{dense}            & store the tensor densely \\
    \texttt{csr}/\texttt{csc} & compress one dimension, in row or column order \\
    \texttt{dcsr}             & compress both dimensions \\
    \texttt{coo}              & store the nonzeros as a coordinate list \\
    \texttt{custom}           & set the storage level of each dimension directly \\
    \midrule
    \texttt{cut\_after\_op}   & end the current fused kernel after this operation \\
    \texttt{extend\_across}   & fuse the operations on both sides of this boundary into one kernel \\
    \midrule
    \texttt{order}            & fix the order in which the loops are iterated \\
    \texttt{par factor}       & set the parallelization factor at a stream level \\
    \texttt{stream shape}     & set the vector width of a stream \\
    \texttt{block-sparse}     & iterate in blocks to exploit block-sparsity \\
    \bottomrule
  \end{tabularx}
\end{table}




\subsection{Ensuring Forward Progress Through Loop Transformations}
\label{sec:instantiations-structure}

\begin{wrapfigure}{r}{0.42\linewidth}
\centering
\resizebox{\linewidth}{!}{%
\begin{tikzpicture}[
  >=stealth,
  nest/.style    ={draw, rounded corners, inner sep=2.5pt, font=\ttfamily\scriptsize},
  merged/.style  ={nest, draw=solarized-magenta, line width=0.8pt},
  edit/.style    ={->, solarized-blue, font=\scriptsize},
  withheld/.style={->, dashed, solarized-base1, font=\scriptsize},
]
  \node[nest]                               (a) {x\ \ y};
  \node[nest, below left=9mm and 7mm of a]  (b) {xo\ xi\ y};
  \node[nest, below right=9mm and 7mm of a] (c) {x\ yo\ yi};
  \node[merged, below=20mm of a]            (d) {xo\ xi\ yo\ yi};
  \draw[edit] (a) -- (b) node[midway, fill=white, inner sep=0.5pt] {split x};
  \draw[edit] (a) -- (c) node[midway, fill=white, inner sep=0.5pt] {split y};
  \draw[edit] (b) -- (d) node[midway, fill=white, inner sep=0.5pt] {split y};
  \draw[edit] (c) -- (d) node[midway, fill=white, inner sep=0.5pt] {split x};
  \draw[withheld] (b) to[bend left=30]
        node[midway, fill=white, inner sep=0.5pt] {fuse} (a);
\end{tikzpicture}}
\caption{
Part of the frontier of nest \texttt{x\,y}. 
Splitting \texttt{x} and then \texttt{y} reaches the same nest as splitting \texttt{y} and then \texttt{x}. 
Fusing \texttt{xo} and \texttt{xi} would return to \texttt{x\,y}, so it is not offered.
}
\label{fig:loopnest-frontier}
\end{wrapfigure}
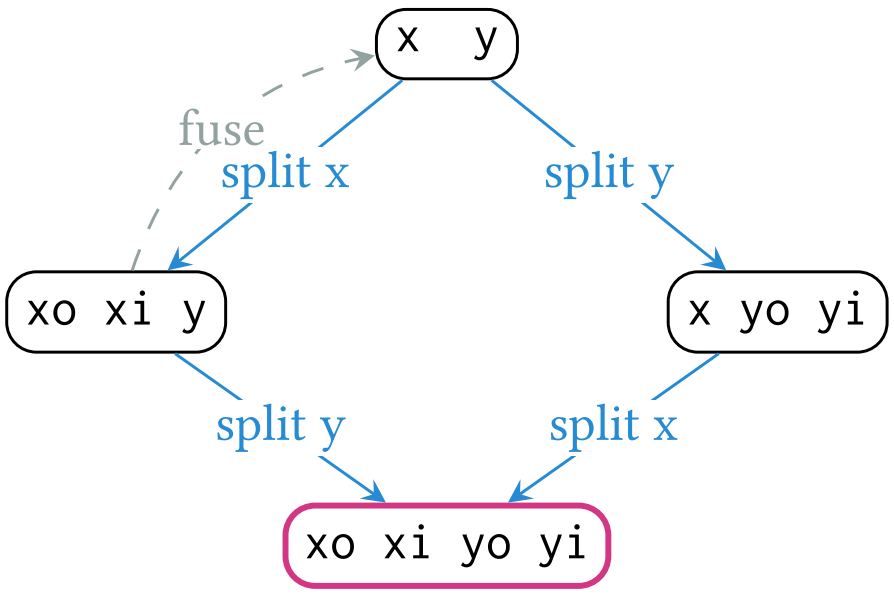

PBN requires that steps never loop back to an expression we have already explored~\cite{lubin:honeybee}, but scheduling languages commonly include operations that can undo one another.
We guarantee forward progress by withholding any step whose result is equivalent to a state already seen in the current interaction. 
We present this approach for Halide, where every function that is not inlined has its own loop nest.
The same technique applies to other loop-based scheduling languages.


Because different sequences of scheduling operations can produce the same loop nest, we detect revisited nests by comparing canonical forms computed with an e-graph~\cite{willsey:egg}. 
\Cref{fig:loopnest-frontier} shows some of the edits available from the nest \code{x\,y}.
Splitting \code{x} and then \code{y} reaches the same nest as splitting \code{y} and then \code{x} (highlighted), even though the steps are taken in a different order.
Fusing \code{xo} and \code{xi} produces a syntactically new loop that is equal to \code{x}, so the dashed edge would return to \code{x\,y} and is withheld.
To compute a canonical form, we represent each loop by the sequence of splits and fuses that derived it from the function's original dimensions.
Two nests are equivalent when they contain the same canonical loops in the same order.

We define the \emph{frontier} of a loop nest $N$ as the set of edits the step provider offers for $N$, and its \emph{path} as the sequence of loop nests the interaction passed through to reach $N$, including $N$ itself.
\Cref{alg:edits} computes the frontier.
It collects every split of a loop by a factor of its extent and every fuse or swap of two adjacent loops, and keeps only those whose result is not equivalent to a nest on the path.
To check equivalence, we add each candidate's resulting nest to the e-graph, which already contains every nest on the path, and compare its canonical form against theirs.
Importantly, withholding edits that return to an earlier nest never makes any nest unreachable.
If a sequence of splits, fuses, and swaps revisits a nest, we can cut out the detour between the two visits and reach the same nest without it.

\begin{algorithm}[t]
\caption{Splits, fuses, and swaps offered for a loop nest $N$}
\label{alg:edits}
\begin{algorithmic}[1]
\Function{Edits}{$N$, \textit{path}}
  \State $E \gets \varnothing$
  \ForAll{loops $\ell \in N$ and factors $f$ with $f \mid \mathrm{extent}(\ell)$}
    \State $E \gets E \cup \{\,\textsc{split}(\ell, f)\,\}$
  \EndFor
  \ForAll{adjacent pairs $(\ell, \ell') \in N$}
    \State $E \gets E \cup \{\,\textsc{fuse}(\ell, \ell')\,,\textsc{swap}(\ell, \ell')\}$
  \EndFor
  \State \Return $\{\, e \in E : \textsc{canon}(e(N)) \notin \textsc{canon}(\textit{path}) \,\}$
  \Comment{\textsc{canon} = e-graph canonical form}
\EndFunction
\end{algorithmic}
\end{algorithm}

Determining the final loop order does not require equivalence checking.
Once a nest's splits are fixed, our instantiation builds its final order from the outermost loop inward.
Each step either places one remaining loop next or fuses two and places the result next; a final step closes the order.
Because each step removes at least one remaining loop, ordering never revisits an earlier state.
Every permutation of the loop nest remains reachable, preserving \CStrongCompleteness{}.

Beyond Halide, this approach can be adapted to other loop-based scheduling languages, including TVM~\cite{chen:tvm}, Tiramisu~\cite{baghdadi:tiramisu}, and several systems surveyed by Hall et al.~\cite{hall:taxonomy}.
In FuseFlow, as with the final loop order, \name{} offers the choice of whether to fuse across each boundary between operations in a single pass from left to right.
Because a decided boundary is never revisited, an interaction can never return to an earlier fusion plan.


\subsection{Defining Schedule Validity}
\label{sec:instantiations-validity}
Not every scheduling operation can be applied in every context, since an operation may be ruled out by the program's semantics, the target hardware, or earlier decisions. 
To offer a step, we need to know whether the resulting sketch can still reach a valid schedule.
Checking this directly would require searching over every way to fill the remaining holes.
Instead, we observe that every scheduling decision has a neutral option that adds no constraints, so a sketch can be completed when filling each remaining hole with its neutral option yields a valid schedule.


\paragraph{FuseFlow}
In FuseFlow, sparse format selections conflict with the loop order.
A sparse format describes a tensor's storage as a hierarchy of levels, in which a compressed level can only be traversed after the levels above it~\cite{chou:formats}. 
Thus, each sparse format constrains the order in which the tensor's dimensions are iterated.
FuseFlow collects these requirements in a \emph{partial order graph} (POG)~\cite{lacouture:fuseflow}, with an edge from index $u$ to index $v$ whenever some tensor requires $u$ to be iterated before $v$.
The legal loop orders are the topological orders of the POG, so a sparse format assignment is valid as long as at least one such order exists (i.e., the POG is acyclic). 
The neutral format is dense, which adds no edges.
To check a candidate format, we combine its edges with those of the formats already chosen, treat every unassigned tensor as dense, and check that the POG is acyclic.
In \Cref{fig:ff-anno}, storing $A$ as \code{csr} requires traversing $i$ before $j$, which closes a cycle with the edges from $B$, $C$, and $D$, so \name{} offers only \code{dense} and \code{csc}.

We implement this check with the SMT solver Z3~\cite{demoura:z3}. 
Although we could check whether the POG is acyclic directly, an SMT encoding lets us add other validity conditions as formulas rather than reimplement the compiler's validation logic inside the step provider.
We give each loop index $u$ an integer rank $r_u$ and assert $r_u < r_v$ for each POG edge $u \to v$.
The query is satisfiable when some loop order respects every edge.
For \code{csr} on $A$ (\Cref{fig:ff-anno}), the constraints require $r_i < r_j < r_k < r_l < r_i$, which is unsatisfiable.
Once every tensor has a format, \name{} offers the POG's topological orders as \code{order} steps, as shown in \Cref{fig:ff-order}.

\begin{figure}[tbp]
\begin{lstlisting}[style=menu,escapechar=!]
// D(i,l) = A(i,j) * B(j,k) * C(k,l)
Working schedule:
  { A: ?,  B: csr,  C: csr,  D: csc }        // choosing A's format

Possible next steps:
  ? !$\mapsto$! A : dense
  ? !$\mapsto$! A : csc
  // not offered: csr, dcsr, coo
\end{lstlisting}
\caption{Choosing a sparse format for $A$, with the other tensors already annotated. 
\name{} offers only \texttt{dense} and \texttt{csc} since \texttt{csr}, \texttt{dcsr}, and \texttt{coo} create a cycle in the POG, leaving no legal loop order.}
\label{fig:ff-anno}
\end{figure}

\begin{figure}[t]
\begin{lstlisting}[style=menu,escapechar=!]
// D(i,l) = A(i,j) * B(j,k) * C(k,l)
Working schedule:
  { order: ?,  par factor: ...,  stream shape: ... }

Possible next steps:
  ? !$\mapsto$! order i j k l                  !$\leftarrow$! taco, scorch
  ? !$\mapsto$! order i j l k
  ? !$\mapsto$! order i k j l
  ...
\end{lstlisting}
\caption{Choosing the dataflow loop order when $A$ is stored with $i$ outermost and $B$, $C$, and $D$ are dense. 
Every legal order places $i$ before $j$.
\name{} offers the twelve legal orders and does not offer the twelve that place $j$ before $i$. 
TACO~\cite{kjolstad:taco-sched} and Scorch~\cite{yan:scorch} both choose $i\,j\,k\,l$, so their hints mark the corresponding step.}
\label{fig:ff-order}
\end{figure}

\paragraph{Halide}
Our Halide instantiation does not need a solver.
By ordering scheduling decisions so that consumers are scheduled before producers, we ensure that every condition on a candidate step involves only decisions that have already been made.
For instance, where a producer can be computed depends on its consumers' loops, so scheduling consumers first means those loops are fixed by the time its placement is decided.

\section{Evaluation}
\label{sec:eval}

We evaluate \name{} as a tool for authoring compiler schedules in Halide and FuseFlow by considering the following three research questions:
\begin{itemize}
    \item \textbf{RQ1}: Can schedules authored with \name{} match or exceed those produced by existing scheduling approaches, i.e., autoschedulers and experts?
    \item \textbf{RQ2}: How does the cost of interactive synthesis scale with the size of the input program?
    \item \textbf{RQ3}: To what extent does \name{} reduce the computational effort required to author high-performance schedules?
\end{itemize}

\subsection{Methodology}
All experiments were run on a workstation with an Intel Core Ultra 9 285K (8 performance and 16 efficiency cores, up to 5.1\,GHz, AVX2, 36\,MB L3 cache) and 64\,GB of RAM. 
Appendix~\ref{app:benchmarks} provides the complete benchmark list, including input sizes, datasets, and configurations.

\paragraph{Halide.}
Halide experiments are run on Ubuntu 24.04 with Halide 21.0.0.
We evaluate 16 applications from the Halide repository's application suite~\cite{halide:repo}, spanning image processing, computational photography, and deep learning, and ranging from 4 to 268 functions (\Cref{tab:benchmarks}).
We use the input and output size estimates from the application code and time every schedule at those sizes. 
We compare \name{} against the three autoschedulers shipped with Halide, Mullapudi2016~\cite{mullapudi:halide-auto}, Li2018~\cite{li:differentiable-halide}, and Adams2019~\cite{adams:learning} for CPU code only\footnote{Because we do not target GPU kernels in this work, we do not compare against Anderson2021~\cite{anderson:halide-gpu}.}.
If one exists, we also compare against the expert-written schedule from the Halide repository. 
Every timing process is pinned to the eight performance cores, and both Halide's thread pool and the autoschedulers' parallelism parameter are set to eight.
For all benchmarks, each schedule is compiled and run once to warm up, then timed in ten rounds of 15 runs and we report the minimum over the 150 runs.

\paragraph{FuseFlow.}
For FuseFlow~\cite{lacouture:fuseflow}, we target the Comal cycle-accurate simulator~\cite{zhang:dam}.
All FuseFlow experiments compile with LLVM 22.0.0 and run in the simulator inside a Docker container, with HBM simulation disabled.
We evaluate the four sparse deep-learning models from the FuseFlow evaluation. 
In total we evaluate 11 model--dataset configurations (\Cref{tab:benchmarks}).
We compare \name{} against FuseFlow's unfused, partially fused, and fully fused configurations and two configurations, without simulator feedback, of the AutoSparse autoscheduler~\cite{lacouture:llm-autosparse}.
TACO~\cite{kjolstad:taco-sched} and Scorch~\cite{yan:scorch} also provide loop-order hints during an interaction. 
We report Comal cycle counts, which are deterministic and independent of the host.

\subsection{RQ1 Results: Schedules Authored with \name{} are Performant}
\label{sec:eval-rq1}

RQ1 asks whether \name{}'s scheduling interface can express performant schedules compared to those produced by autoschedulers or written by experts. 
We author schedules interactively with \name{}, guided by hints from existing schedules and our own knowledge of the scheduling space. 
Our primary objective is to match or exceed the fastest evaluated autoscheduler for each application, and we generally stop refining a schedule once reaching this target. 
\Cref{fig:rq1-geomeans} summarizes the geometric-mean speedup of each scheduling method for Halide and FuseFlow. 

\begin{figure}[t]
  \centering
  \includegraphics[width=\linewidth]{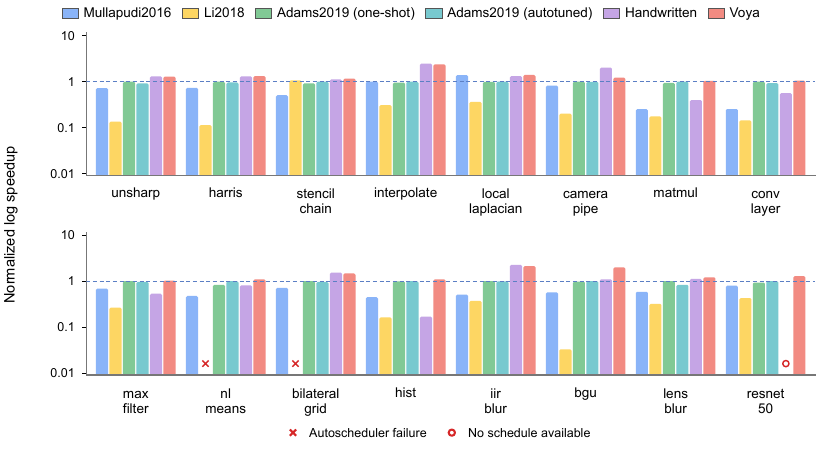}
  \caption{
  Log speedup of each Halide schedule normalized to the best Adams2019 configuration (one-shot vs. autotuned), where the dashed line marks the baseline. 
  Legend entries follow bars within each benchmark group from left-to-right.
  Li2018's generated schedules fail with allocation-size errors on two kernels.
  \code{resnet\_50} has no expert-written schedule in the Halide repo~\cite{halide:repo}.
  Overall, schedules authored with \name{} achieve a geometric-mean speedup of 1.04$\times$ over the fastest existing schedule and a 1.34$\times$ geometric-mean speedup over the fastest Adams2019 schedules.
  }
  \label{fig:halide-speedup}
\end{figure}

\paragraph{Halide.}
\Cref{fig:halide-speedup} shows, for each Halide application, the speedup of each method's schedule relative to the faster of the schedules produced by two Adams2019 configurations.
Both configurations use the pretrained cost-model weights distributed with Halide.
The \textit{one-shot} configuration produces a schedule through a single beam search.
The \textit{autotuned} configuration benchmarks 32 candidates (one from beam search and 31 from randomized searches guided by the cost model), selects the fastest, and remeasures that schedule for the reported comparison.
Relative to the fastest Adams2019 schedules, \name{} schedules achieve a geometric-mean speedup of 1.34$\times$, ranging from 1.04$\times$ on \code{max\_filter} to 2.40$\times$ on \code{interpolate}. 
Compared to the fastest evaluated autoscheduler for each application, \name{} achieves a geometric-mean speedup of 1.31$\times$. 

Across all 16 Halide applications, schedules authored with \name{} achieve a geometric-mean speedup of 1.04$\times$ over the fastest existing schedule, whether expert-written or generated by an autoscheduler, with a range of 0.60$\times$--1.81$\times$.
We treat expert schedules as secondary targets because our library only implements the subset of Halide's scheduling operations shown in \Cref{tab:halide-steps}.
\name{} schedules underperform compared to the expert-written schedules on five applications. 
The largest gap is on \code{camera\_pipe}, where the expert schedule uses directives such as \code{compute\_with}, \code{fold\_storage}, and \code{prefetch}, which our library does not yet include. 
The largest improvement is on \code{bgu}, where we use an existing autoscheduler-generated schedule as a starting point and adapt it through \name{} to outperform the expert-written baseline.

\begin{figure}[t]
  \centering
  \includegraphics[width=\linewidth]{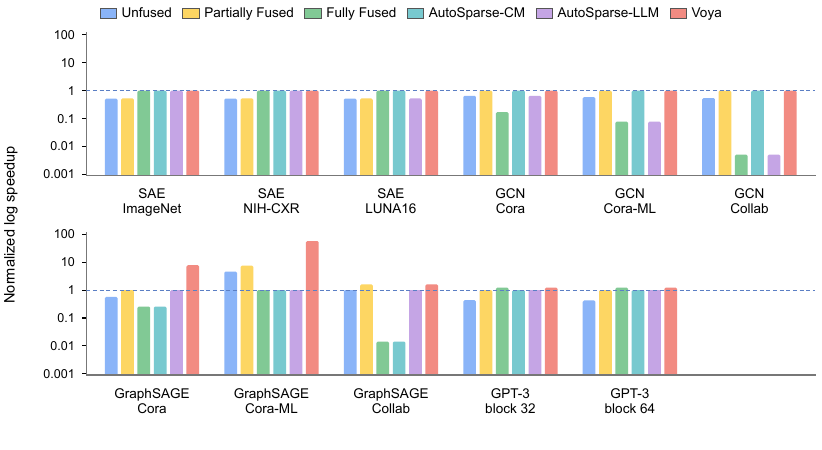}
  \caption{
  Log speedup of each FuseFlow schedule over AutoSparse, in Comal cycles. 
  Legend entries follow bars within each benchmark group from left-to-right.
  The AutoSparse baseline is the faster of AutoSparse-CM and AutoSparse-LLM, denoted by the dashed line.
  Overall, schedules authored with \name{} achieve a geometric-mean speedup of 1.45$\times$ over the fastest existing schedule and a 1.88$\times$ geometric-mean speedup over the fastest autoscheduler (see \Cref{fig:rq1-geomeans} bottom). 
  }
  \label{fig:fuseflow-speedup}
\end{figure}

\paragraph{FuseFlow.}
\Cref{fig:fuseflow-speedup} shows the schedule speedup of  cycle counts normalized to the fastest AutoSparse configuration, as with Halide. 
\textit{AutoSparse-CM} ranks fusion plans using its analytical cost model, and \textit{AutoSparse-LLM} asks an LLM (Claude Opus~5.5) to propose ten candidate plans in one round  and re-ranks the compilable ones with the same cost model.
Relative to the fastest autoscheduler, \name{} schedules achieve a geometric-mean speedup of 1.88$\times$, ranging from 1$\times$ to 56.8$\times$.
The largest improvement over AutoSparse is on GraphSAGE with Cora-ML, where both AutoSparse configurations select full fusion. 
With \name{}, we instead adopt the existing partially fused plan and add eight-way parallelization, yielding a 56.8$\times$ speedup over AutoSparse and a 7.55$\times$ speedup over the partially fused baseline.
Across all 11 configurations, \name{} schedules ultimately achieve a geometric-mean speedup of 1.45$\times$ over the fastest existing schedule, with a range of 1$\times$--7.74$\times$. 

\subsection{RQ2 Results: \name{} Scales to Large Programs}
\label{sec:eval-rq2}
Since synthesis scaling may be a concern, RQ2 asks how the computational cost of interactive synthesis scales with program size. 
Using the best schedules from \Cref{sec:eval-rq1} for each target application, we use a deterministic step decider to recreate them. During this process, we measure the number of rounds, the number of steps provided at each round, and the time to compute the steps at each round. 
Each interaction is repeated three times, with the synthesizer pinned to one CPU core, and we report the medians across target schedules and repetitions. 

\begin{figure*}[t]
  \centering
  \begin{minipage}[t]{1.35in}
    \vspace{0pt}
    \centering
    \includegraphics[width=\linewidth]{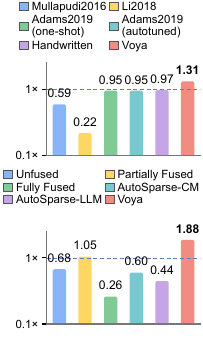}
    \caption{
    Geometric-mean speedup (Halide top, FuseFlow bottom) relative to best prior autoscheduler. 
    }
    \label{fig:rq1-geomeans}
  \end{minipage}\hfill%
  \begin{minipage}[t]{4.05in}
    \vspace{0pt}
    \centering
    \includegraphics[scale=0.9]{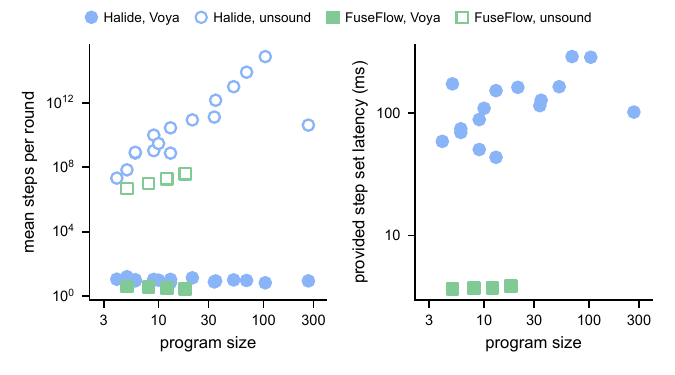}
    \caption{
    Interactive synthesis cost as program size increases.
    The left plot shows the mean number of steps per round for \name{} and the unsound baseline.
    The right plot shows the median time to compute \name{}'s provided step set.
    }
    \label{fig:rq2}
  \end{minipage}
\end{figure*}

We compare \name{} against a baseline step provider that enumerates candidate steps using type information but ignores scheduling constraints, thereby sacrificing \CStrongSoundness{}.
Because this baseline exhausts the 16\,GB memory limit on every benchmark, we count its candidates analytically from combinations of scheduling operations and parameter values, without constructing them.
\Cref{fig:rq2} plots both providers' mean candidate counts per round and the median time to compute \name{}'s steps against program size.


\paragraph{Halide.}
Across the 16 applications, the average number of \name{} provided steps per round is between 6.7 and 16.0, with no individual round offering more than 67 steps. 
In contrast, the mean candidate count of the baseline ranges from 2.13$\times10^7$ on \code{matmul} to 7.45$\times10^{14}$ on \code{local\_laplacian}. 
The median time to compute \name{}'s provided step set ranges from 44\,ms on \code{harris} to 290\,ms on \code{lens\_blur} and no individual round exceeds 0.92\,s.

\paragraph{FuseFlow.}
Across the 11 configurations, \name{} provides an average of 2.9--4.0 steps per round, with no individual round offering more than 9.
The baseline's mean step count ranges from 4.82$\times10^6$ on SAE to 3.89$\times10^7$ on GPT-3.
Computing \name{}'s provided steps takes a median 4\,ms per round, and complete interactions require 9--20 rounds and take 0.1--0.2\,s.

\subsection{RQ3 Results: LLM Agents with \name{} Require Fewer Tokens for FuseFlow Scheduling}
\label{sec:eval-rq3}
Although \name{} is primarily intended for users seeking direct control over at least some scheduling decisions, LLMs are increasingly writing and optimizing programs.
Therefore, RQ3 asks whether \name{} reduces the computational effort of generating performant Halide and FuseFlow schedules when decisions are made exclusively by AI.
We compare LLM agents writing schedules directly with those using \name{} interactively.
We measure effort as cumulative output tokens. 
At each token count, we report the performance of the best correct schedule generated so far, normalized to the best performance achieved by any approach during the trial; zero indicates no correct schedule yet.
We evaluate Claude Code (Anthropic) Opus~5.5 for FuseFlow and Halide, using a custom harness with one trial per model, benchmark, and interface.
Each trial uses the \texttt{high} effort setting, with a two-hour wall-clock limit and an output budget of 64,000 tokens for FuseFlow or 16,000 for Halide. 
We evaluate \name{} with and without hints for both FuseFlow and Halide.

\begin{figure*}[t]
  \centering
  \includegraphics[width=\textwidth]{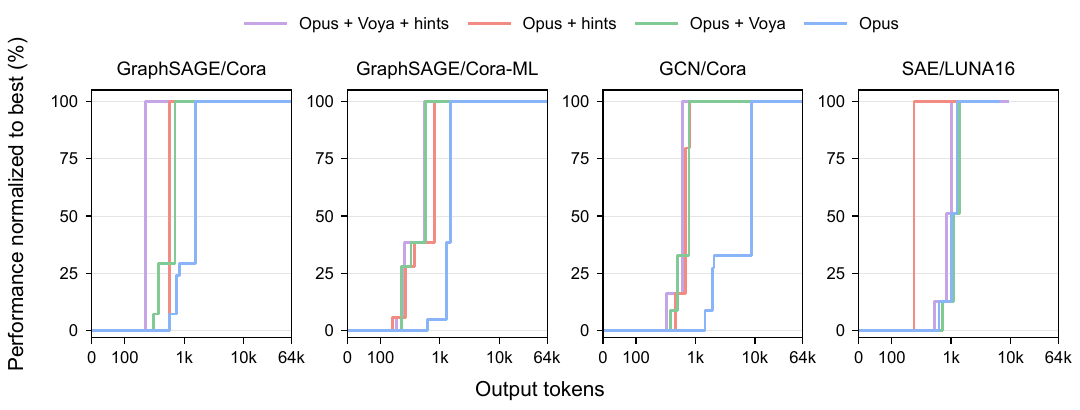}
  \caption{
  FuseFlow schedule performance versus output tokens. 
  Curves show best performance per trial as a percentage of the performance of the best schedule generated by the two approaches. 
  Output tokens include thinking. 
  Horizontal axis uses symlog scale. 
  For FuseFlow, agents with \name{} require fewer tokens to reach 100\% performance on average.
  }
  \label{fig:rq3-agents-fuseflow}
\end{figure*}

\paragraph{FuseFlow}
We evaluate nine of the eleven workload--input configurations from \Cref{sec:eval-rq1} and omit the GPT-3 configurations because their simulations exceed an hour, preventing iteration within the two-hour budget.
\Cref{fig:rq3-agents-fuseflow} plots schedule performance against token usage for four of these nine configurations, with the full plot in \Cref{fig:rq3-agents-fuseflow-app} in Appendix~\ref{app:rq3-full}.
\name{} without hints reduces token usage by a geometric mean of 2.30$\times$ across all configurations for Opus~5.5. 
The savings reach 11.2$\times$ on GCN/Cora, while the worst case is SAE/LUNA16, where \name{} uses 1.1$\times$ more tokens.
For larger SAE kernels, costly schedule evaluations consumed much of the two-hour budget and limited search iterations, leaving agents well short of their 64,000-token allowance and making differences in token efficiency harder to discern. 
With hints enabled, \name{} reduced output-token usage by a geometric mean of 1.37$\times$. 
These results suggest that hints help agents directly identify promising schedules, reducing the advantage of interactive synthesis.

Across FuseFlow experiments without hints, 9.3\% of schedules authored without \name{} were rejected at compile time, compared to none with \name{}. 
With hints, 11.3\% of schedules authored without \name{} directly failed to compile, compared to none with \name{}.  
These results show that \name{} can reduce schedule authoring effort even when the agent ultimately reaches the same performance both with and without \name{}. 
They also highlight the value of \CStrongSoundness{} when validity depends on interdependent scheduling decisions.

\begin{figure*}[t]
  \centering
  \includegraphics[width=\textwidth]{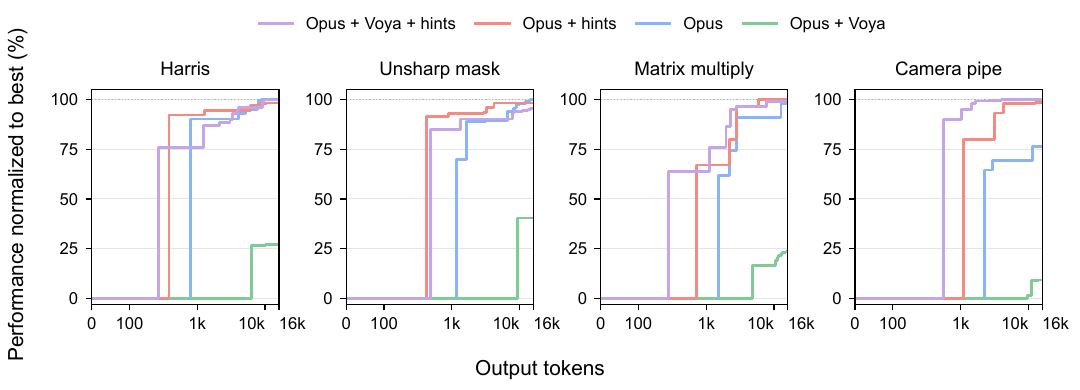}
  \caption{
  Halide schedule performance versus output tokens. 
  Curves show best performance per trial as a percentage of the performance of the best schedule generated by any of the four approaches. 
  Horizontal axis uses symlog scale. 
  For Halide, agents with \name{} but without hints require more tokens and reach worse schedules. 
  With hints, they reach comparable performance and require fewer tokens on average.
  }
  \label{fig:rq3-agents-halide}
\end{figure*}

\paragraph{Halide}
Similarly, \Cref{fig:rq3-agents-halide} shows schedule performance as LLM agents consume more tokens for four of the Halide benchmarks in \Cref{sec:eval-rq1}. 
Across the four Halide benchmarks, agents using \name{} without hints spend more tokens than agents writing schedules directly. 
Further, agents using \name{} without hints write schedules that are 4.24$\times$ slower by geometric mean. 
With hints, both interfaces reach 95\% of the best performance on all four kernels, and \name{} reduces the required tokens by a geometric mean of 1.16$\times$. 
Unlike FuseFlow, agents write valid Halide schedules reliably. 
We hypothesize that models have encountered more Halide schedules than FuseFlow schedules during training, making them more effective at optimizing Halide. 
With hints, we suspect that both interfaces instead start from or adapt the provided reference schedules, which have already been carefully optimized. However, note that these Opus + \name{} trends are a reversal of the pattern in FuseFlow, where \name{} without any hints still outperformed the direct agent without \name{}.

\paragraph{Case Study: Enabling Scheduling with a Probabilistic Decision Model}
To explore lower inference costs, we evaluated Jev~\cite{almeida:jev}, a non-generative probabilistic decision model that guarantees type-safe outputs, on GraphSAGE. We chose GraphSAGE because Opus+\name{} dominated the most in our FuseFlow experiments above. 
Such classifiers cannot construct schedules at all without \name{}. 
With a 600\,s search budget (\Cref{tab:graphsage-jev}), Jev+\name{} finds schedules within one cycle of Opus+\name{} on Cora and Cora-ML, at approximately 10$\times$ and 31$\times$ lower cumulative model cost at each run’s best-schedule discovery. 
On Collab, Jev+\name{} finds a schedule with 31\% more execution cycles at approximately $29\times$ lower model cost to discovery. 
This case study shows \name{} enables probabilistic decision models to provide reasonably strong schedules at substantially lower inference cost.

\begin{table}[t]
\centering
\small
\setlength{\tabcolsep}{4pt}
\begin{tabular}{@{}lcccccc@{}}
\toprule
Dataset & \multicolumn{2}{c}{Cycles} & \multicolumn{2}{c}{Time (s)} & \multicolumn{2}{c}{Cost (\$)} \\
\cmidrule(lr){2-3}
\cmidrule(lr){4-5}
\cmidrule(lr){6-7}
& Jev+\name{} & Opus+\name{} & Jev+\name{} & Opus+\name{} & Jev+\name{} & Opus+\name{} \\
\midrule
Cora    & 4,518 & 4,517 & 184.9 & 49.1 & 0.0220 & 0.2223 \\
Cora-ML & 4,118 & 4,117 & 127.7 & 82.5 & 0.0137 & 0.4226 \\
Collab  & 6,862 & 5,237 & 129.8 & 87.0 & 0.0148 & 0.4230 \\
\bottomrule
\end{tabular}
\caption{GraphSAGE scheduling with a 600\,s search budget.
Cycles measure the final best schedule's execution time.
Time and cumulative model cost are measured as time and cost to reach the best schedule.
}
\label{tab:graphsage-jev}
\end{table}

\section{Related Work}
\name{} enables programmers to author schedules for Halide and FuseFlow interactively, composing their own decisions with those from existing schedules while maintaining validity and reachability guarantees.
We discuss related work on interactive optimization, autoscheduling, formal guarantees for scheduling languages, and LLM-driven optimization. 

\subsection{Interactive and Guided Compiler Optimization}
We are not the first to argue for interactivity in optimizing compilers~\cite{kennedy:parascope, liao:suif-explorer, zinenko:clint, bagneres:blackbox, papenhausen:pumav}. 
\textit{Guided equality saturation}~\cite{koehler:guided-eqsat} queries the user for intermediate goals at challenging points in an equality saturation workload; its goal is to be mostly automated while occasionally involving the user.
Because a goal may not be reachable, a user's input can lead to no solution, so guided equality saturation does not offer \CStrongSoundness{}.
By contrast, \name{} does offer \CStrongSoundness{}, but at a finer granularity. 
Where guided equality saturation lets users state a goal and leaves the intermediate rewrites to the solver, \name{} asks users to make every scheduling decision explicitly. 

The \textit{guided optimization} system Roly-poly~\cite{ikarashi:roly-poly} enables users to follow a fixed-order workflow of i)~choosing function scheduling locations and ii)~tiling parameters for a Halide program.
It does not enable users to reach every possible schedule in the scheduling language, but, unlike guided equality saturation, it does guarantee that every option it shows is valid.
Roly-poly users produced significantly improved schedules, but they were still outperformed by the autoscheduler.
Like Roly-poly, \name{} shows only valid options, but it also offers \CStrongCompleteness{} with respect to the scheduling directives it supports.
Moreover, where Roly-poly ranks options with a single cost model, \name{} can draw hints from any number of existing schedules whose hints can be composed (e.g., by averaging) to rank the offered steps.  
Unlike Roly-poly, however, \name{} gives no performance estimates and has not been evaluated with human users.

\subsection{Autoscheduling}
Autoschedulers produce schedules automatically for both dense~\citep{mullapudi:halide-auto, li:differentiable-halide, adams:learning, anderson:halide-gpu, chen:autotvm, zheng:ansor, shao:metaschedule, haj-ali:protuner} and sparse~\citep{won:waco, dias:sparseauto, dias:sparselnr, ahrens:auto, yan:scorch, lacouture:llm-autosparse} programs.
As discussed in \Cref{sec:pbn}, any autoscheduler can serve as a step decider, and we build directly on this body of work by ingesting several autoschedulers into \name{}~\cite{lacouture:llm-autosparse, yan:scorch, kjolstad:taco-sched, mullapudi:halide-auto, li:differentiable-halide, adams:learning}.
As such, the rest of this section focuses on other autoschedulers.

Autoschedulers ensure validity in one of two ways.
Some restrict the search space so that every point in it is valid, giving up \CStrongCompleteness{}, as Halide's autoschedulers do~\citep{adams:learning, haj-ali:protuner}.
Others search a richer space and discard invalid candidates after generating them, giving up \CStrongSoundness{}~\citep{zheng:ansor, shao:metaschedule}. 
MetaSchedule~\cite{shao:metaschedule} records each candidate as a trace of sampling decisions, much like \name{}'s traces and hints, but its search mutates these decisions and relies on a validator to reject the invalid traces that result, whereas \name{} never offers an invalid step.
MetaSchedule also offers composition, combining transformation modules into a search space in which a single search algorithm makes every decision, whereas \name{} composes the decisions themselves.
Telamon~\citep{beaugnon:telamon} lets decisions be fixed in any order, but its constraint propagation can miss incompatibilities, so its search can reach dead ends.
Sparse autoschedulers each decide a different part of a schedule~\citep{lacouture:llm-autosparse, dias:sparseauto, dias:sparselnr, ahrens:auto}.
We hypothesize that any of these search strategies could serve as a step decider over \name{}'s offered steps, yielding autoschedulers whose outputs are valid by construction.

\subsection{Formal Guarantees for Scheduling Languages}
Many prior works guarantee that scheduling cannot change what a program computes.
\citet{reinking:halide-semantics} formalized Halide's semantics and showed that every valid schedule yields a memory-safe program that computes the same result as the algorithm.
Other systems instead analyze the program being scheduled each time a transformation is applied, rejecting any that would change the program's result~\cite{feng:tensorir, ikarashi:exo, ikarashi:exo2}.
EXO 2~\cite{ikarashi:exo2} extends this guarantee to user-defined scheduling operators by providing a core set of primitives that always preserve the program's behavior.
In doing so, it also offers a notion of composition, in which new scheduling operators are built from these primitives.
By contrast, \name{}'s composition concerns the scheduling decisions themselves and guarantees that any sequence of decisions ends in a valid schedule.

Another line of work expresses scheduling as rewrites whose correctness is established once per rule rather than per program~\cite{steuwer:lift, hagedorn:rise-elevate, arora:tensorright}.
ATL further provides mechanized proofs of correctness for each scheduling rewrite and for lowering to imperative code, and has been extended to sparse tensor formats~\cite{liu:atl-popl22, liu:atl-pldi24, liu:atl-pldi26}.
Like these systems, \name{} offers formal guarantees, but where they ensure that a schedule or rewrite preserves the meaning of the program, \CStrongSoundness{} and \CStrongCompleteness{} instead concern how schedules are constructed.
We believe \name{}'s guarantees are complementary to all of these.
While \name{} ensures that every interaction ends in a valid schedule, semantic guarantees ensure that any valid schedule is correct.

\subsection{Language Models for Code Optimization}
Language models are increasingly used to generate and transform code, with varying degrees of freedom. We consider prior work on LLMs and agents that do not author schedules directly.
At one extreme, models generate optimized kernels directly, with correctness established by tests or by verifiers applied afterward~\citep{ouyang:kernelbench, costa:ai-compiler, zhang:ptxbench}.
\citet{costa:ai-compiler} lower Triton kernels directly to PTX, outperforming the Triton compiler by applying transformations its pipeline does not express, but find that restricting generation to code their verifier supports erases most of this advantage.
PTXBench evaluates agents that generate CUDA kernels with PTX and refine them with feedback~\citep{zhang:ptxbench}.
Others propose transformations to an existing compiler, which checks their legality and measures their effect~\cite{merouani:compilot, hong:autocomp}.
\name{} constrains the model further, offering only steps that can reach a valid schedule, so every schedule it reaches inherits the compiler's correctness guarantees.
Unlike direct generation, \name{} can reach only schedules the scheduling language can express.

\section{Conclusion}
We present a general approach for viewing schedules as sketches in domain-specific compilers while providing validity and reachability guarantees.
Our strategy enables the composition of scheduling decisions from programmers, autoschedulers, and preexisting handwritten schedules, thereby addressing the limitations of relying on manual scheduling or autoscheduling alone to meet programmers' objectives.
Building upon these insights, we implement \name{}, an interactive scheduling tool for Halide and FuseFlow.
Our instantiations demonstrate that the method applies to different domain-specific languages, varied workloads, and both general-purpose CPUs and specialized accelerators.
Our results show that \name{} enables programmers to construct high-performance schedules and reduces the number of tokens LLMs need to produce such schedules for systems with limited autoscheduler support and few manually tuned schedules.
We hope programmers use \name{} for schedule authoring and compiler developers apply our approach to other scheduling languages. 
We also hope this work inspires research into new approaches to schedule authoring and provides a foundation for interactive scheduling with strong guarantees.



\section*{Acknowledgments}
We thank Nikhil Agarwal, Matteo Cericola, Sun A Cho, Malena Cochrane, Rena Feng, Mitchell Fream, Humza Ikram, Ethan Lu, Stephen Mell, Nathan Serafin, Guy Wilks, and Rohan Yadav for their feedback on this paper.
We also thank Andrew Adams, Limin Jia, Jaehyun Lim, AJ Root, Gaurav Shah, and Krish Sundarajah for their insights and discussions on this work.
This work was supported in part by DARPA under the Machine learning and Optimization-guided Compilers for Heterogeneous Architectures (MOCHA) program (Award No. HR00112520038), the NSF under the Cyberinfrastructure Systems for Scientific Innovation (CSSI) program (Award No. 2609118), a Google ML and Systems Junior Faculty Award, and a grant of access to OpenAI models through the ChatGPT for Academic Researchers program.
Evan Williams was supported in part by the NSF CISE Graduate Fellowship (Award No. 2313998).  
Any opinions, findings, conclusions, or recommendations expressed in this material are those of the authors and do not necessarily reflect the views of the aforementioned funding agencies and organizations.

\bibliographystyle{ACM-Reference-Format}
\bibliography{references}

@article{adams:learning,
  title={Learning to optimize halide with tree search and random programs},
  author={Adams, Andrew and Ma, Karima and Anderson, Luke and Baghdadi, Riyadh and Li, Tzu-Mao and Gharbi, Micha{\"e}l and Steiner, Benoit and Johnson, Steven and Fatahalian, Kayvon and Durand, Fr{\'e}do and others},
  journal={ACM Transactions on Graphics (TOG)},
  volume={38},
  number={4},
  pages={1--12},
  year={2019},
  publisher={ACM New York, NY, USA}
}

@misc{adobe:halide,
  author       = {{Adobe Research}},
  title        = {Inside {Halide}, the Open Source Language Engineers Use to Make Imaging Tools Faster},
  year         = {2025},
  month        = apr,
  day          = {30},
  howpublished = {\url{https://research.adobe.com/news/inside-halide-the-open-source-language/}},
  lastaccessed = {October 23, 2026},
}

@inproceedings{ahrens:auto,
author = {Ahrens, Willow and Kjolstad, Fredrik and Amarasinghe, Saman},
title = {Autoscheduling for sparse tensor algebra with an asymptotic cost model},
year = {2022},
isbn = {9781450392655},
publisher = {Association for Computing Machinery},
address = {New York, NY, USA},
url = {https://doi.org/10.1145/3519939.3523442},
doi = {10.1145/3519939.3523442},
booktitle = {Proceedings of the 43rd ACM SIGPLAN International Conference on Programming Language Design and Implementation},
pages = {269–285},
numpages = {17},
location = {San Diego, CA, USA},
series = {PLDI 2022}
}

@misc{almeida:jev,
  author       = {Almeida, Diogo},
  title        = {Introducing {System One Models} \& {Jev}},
  howpublished = {TypeSafe AI Blog},
  year         = {2026},
  month        = sep,
  url          = {https://typesafe.ai/blog/introducing-system-one-models-and-jev},
}

@inproceedings{anderson:halide-gpu,
author = {Anderson, Luke and Adams, Andrew and Ma, Karima and Li, Tzu-Mao and Jin, Tian and Ragan-Kelley, Jonathan},
title = {Efficient automatic scheduling of imaging and vision pipelines for the GPU},
year = {2021},
issue_date = {October 2021},
publisher = {Association for Computing Machinery},
address = {New York, NY, USA},
volume = {5},
number = {OOPSLA},
url = {https://doi.org/10.1145/3485486},
doi = {10.1145/3485486},
journal = {Proc. ACM Program. Lang.},
month = oct,
articleno = {109},
numpages = {28}
}

@inproceedings{ansel:opentuner,
  title={Opentuner: An extensible framework for program autotuning},
  author={Ansel, Jason and Kamil, Shoaib and Veeramachaneni, Kalyan and Ragan-Kelley, Jonathan and Bosboom, Jeffrey and O'Reilly, Una-May and Amarasinghe, Saman},
  booktitle={Proceedings of the 23rd international conference on Parallel architectures and compilation},
  pages={303--316},
  year={2014}
}

@online{apache:tvm,
  author  = {{The Apache Software Foundation}},
  title   = {The {Apache} Software Foundation Announces {Apache} {TVM} as a Top-Level Project},
  url     = {https://news.apache.org/foundation/entry/the-apache-software-foundation-announces69},
  date    = {2020-11-30},
  urldate = {2026-10-28}
}

@article{arora:tensorright,
author = {Arora, Jai and Lu, Sirui and Jain, Devansh and Xu, Tianfan and Houshmand, Farzin and Phothilimthana, Phitchaya Mangpo and Lesani, Mohsen and Narayanan, Praveen and Murthy, Karthik Srinivasa and Bodik, Rastislav and Sabne, Amit and Mendis, Charith},
title = {TensorRight: Automated Verification of Tensor Graph Rewrites},
year = {2025},
issue_date = {January 2025},
publisher = {Association for Computing Machinery},
address = {New York, NY, USA},
volume = {9},
number = {POPL},
url = {https://doi.org/10.1145/3704865},
doi = {10.1145/3704865},
journal = {Proc. ACM Program. Lang.},
month = jan,
articleno = {29},
numpages = {32}
}

@inproceedings{baghdadi:tiramisu,
  title={Tiramisu: A polyhedral compiler for expressing fast and portable code},
  author={Baghdadi, Riyadh and Ray, Jessica and Romdhane, Malek Ben and Del Sozzo, Emanuele and Akkas, Abdurrahman and Zhang, Yunming and Suriana, Patricia and Kamil, Shoaib and Amarasinghe, Saman},
  booktitle={2019 IEEE/ACM International Symposium on Code Generation and Optimization (CGO)},
  pages={193--205},
  year={2019},
  organization={IEEE}
}

@misc{baghdadi:tiramisu-auto,
      title={A Deep Learning Based Cost Model for Automatic Code Optimization}, 
      author={Riyadh Baghdadi and Massinissa Merouani and Mohamed-Hicham Leghettas and Kamel Abdous and Taha Arbaoui and Karima Benatchba and Saman Amarasinghe},
      year={2021},
      eprint={2104.04955},
      archivePrefix={arXiv},
      primaryClass={cs.PL},
      url={https://arxiv.org/abs/2104.04955}, 
}

@inproceedings{bagneres:blackbox,
author = {Bagn{\`e}res, L{\'e}na{\"i}c and Zinenko, Oleksandr and Huot, St{\'e}phane and Bastoul, C{\'e}dric},
title = {Opening polyhedral compiler's black box},
year = {2016},
isbn = {9781450337786},
publisher = {Association for Computing Machinery},
address = {New York, NY, USA},
url = {https://doi.org/10.1145/2854038.2854048},
doi = {10.1145/2854038.2854048},
booktitle = {Proceedings of the 2016 International Symposium on Code Generation and Optimization},
pages = {128–138},
numpages = {11},
location = {Barcelona, Spain},
series = {CGO '16}
}

@article{bansal:mosaic,
author = {Bansal, Manya and Hsu, Olivia and Olukotun, Kunle and Kjolstad, Fredrik},
title = {Mosaic: An Interoperable Compiler for Tensor Algebra},
year = {2023},
issue_date = {June 2023},
publisher = {Association for Computing Machinery},
address = {New York, NY, USA},
volume = {7},
number = {PLDI},
url = {https://doi.org/10.1145/3591236},
doi = {10.1145/3591236},
journal = {Proc. ACM Program. Lang.},
month = jun,
articleno = {122},
numpages = {26}
}

@inproceedings{beaugnon:telamon,
author = {Beaugnon, Ulysse and Pouille, Antoine and Pouzet, Marc and Pienaar, Jacques and Cohen, Albert},
title = {Optimization space pruning without regrets},
year = {2017},
isbn = {9781450352338},
publisher = {Association for Computing Machinery},
address = {New York, NY, USA},
url = {https://doi.org/10.1145/3033019.3033023},
doi = {10.1145/3033019.3033023},
booktitle = {Proceedings of the 26th International Conference on Compiler Construction},
pages = {34–44},
numpages = {11},
location = {Austin, TX, USA},
series = {CC 2017}
}

@article{chen:allo,
author = {Chen, Hongzheng and Zhang, Niansong and Xiang, Shaojie and Zeng, Zhichen and Dai, Mengjia and Zhang, Zhiru},
title = {Allo: A Programming Model for Composable Accelerator Design},
year = {2024},
issue_date = {June 2024},
publisher = {Association for Computing Machinery},
address = {New York, NY, USA},
volume = {8},
number = {PLDI},
url = {https://doi.org/10.1145/3656401},
doi = {10.1145/3656401},
journal = {Proc. ACM Program. Lang.},
month = jun,
articleno = {171},
numpages = {28}
}

@inproceedings{chen:tvm,
author = {Chen, Tianqi and Moreau, Thierry and Jiang, Ziheng and Zheng, Lianmin and Yan, Eddie and Cowan, Meghan and Shen, Haichen and Wang, Leyuan and Hu, Yuwei and Ceze, Luis and Guestrin, Carlos and Krishnamurthy, Arvind},
title = {TVM: an automated end-to-end optimizing compiler for deep learning},
year = {2018},
isbn = {9781931971478},
publisher = {USENIX Association},
address = {USA},
booktitle = {Proceedings of the 13th USENIX Conference on Operating Systems Design and Implementation},
pages = {579–594},
numpages = {16},
location = {Carlsbad, CA, USA},
series = {OSDI'18}
}

@inproceedings{chen:autotvm,
author = {Chen, Tianqi and Zheng, Lianmin and Yan, Eddie and Jiang, Ziheng and Moreau, Thierry and Ceze, Luis and Guestrin, Carlos and Krishnamurthy, Arvind},
title = {Learning to optimize tensor programs},
year = {2018},
publisher = {Curran Associates Inc.},
address = {Red Hook, NY, USA},
booktitle = {Proceedings of the 32nd International Conference on Neural Information Processing Systems},
pages = {3393–3404},
numpages = {12},
location = {Montr{\'e}al, Canada},
series = {NIPS'18}
}

@inproceedings{chou:formats,
  author = {Chou, Stephen and Kjolstad, Fredrik and Amarasinghe, Saman},
  title = {Format Abstraction for Sparse Tensor Algebra Compilers},
  journal = {Proc. ACM Program. Lang.},
  issue_date = {November 2018},
  volume = {2},
  number = {OOPSLA},
  month = oct,
  year = {2018},
  issn = {2475-1421},
  pages = {123:1--123:30},
  articleno = {123},
  numpages = {30},
  url = {http://doi.acm.org/10.1145/3276493},
  doi = {10.1145/3276493},
  acmid = {3276493},
  publisher = {ACM},
  address = {New York, NY, USA},
}

@misc{costa:ai-compiler,
      title={AI as a Compiler: Compiling Triton kernels without the Triton compiler}, 
      author={François Costa and Charly Castes and Thomas Bourgeat and Azalia Mirhoseini},
      year={2026},
      eprint={2609.36800},
      archivePrefix={arXiv},
      primaryClass={cs.AI},
      url={https://arxiv.org/abs/2609.36800}, 
}

@inproceedings{demoura:z3,
author = {De Moura, Leonardo and Bj{\o}rner, Nikolaj},
title = {Z3: an efficient SMT solver},
year = {2008},
isbn = {3540787992},
publisher = {Springer-Verlag},
address = {Berlin, Heidelberg},
booktitle = {Proceedings of the Theory and Practice of Software, 14th International Conference on Tools and Algorithms for the Construction and Analysis of Systems},
pages = {337–340},
numpages = {4},
location = {Budapest, Hungary},
series = {TACAS'08/ETAPS'08}
}

@article{dias:sparseauto,
  title={SparseAuto: An Auto-scheduler for Sparse Tensor Computations using Recursive Loop Nest Restructuring},
  author={Dias, Adhitha and Anderson, Logan and Sundararajah, Kirshanthan and Pelenitsyn, Artem and Kulkarni, Milind},
  journal={Proceedings of the ACM on Programming Languages},
  volume={8},
  number={OOPSLA2},
  pages={527--556},
  year={2024},
  publisher={ACM New York, NY, USA}
}

@inproceedings{dias:sparselnr,
author = {Dias, Adhitha and Sundararajah, Kirshanthan and Saumya, Charitha and Kulkarni, Milind},
title = {SparseLNR: accelerating sparse tensor computations using loop nest restructuring},
year = {2022},
isbn = {9781450392815},
publisher = {Association for Computing Machinery},
address = {New York, NY, USA},
url = {https://doi.org/10.1145/3524059.3532386},
doi = {10.1145/3524059.3532386},
booktitle = {Proceedings of the 36th ACM International Conference on Supercomputing},
articleno = {15},
numpages = {14},
location = {Virtual Event},
series = {ICS '22}
}

@misc{fegade:cora,
      title={The CoRa Tensor Compiler: Compilation for Ragged Tensors with Minimal Padding}, 
      author={Pratik Fegade and Tianqi Chen and Phillip B. Gibbons and Todd C. Mowry},
      year={2022},
      eprint={2110.10221},
      archivePrefix={arXiv},
      primaryClass={cs.LG},
      url={https://arxiv.org/abs/2110.10221}, 
}

@inproceedings{fegade:cortex,
 author = {Fegade, Pratik and Chen, Tianqi and Gibbons, Phillip and Mowry, Todd},
 booktitle = {Proceedings of Machine Learning and Systems},
 editor = {A. Smola and A. Dimakis and I. Stoica},
 pages = {38--54},
 title = {Cortex: A Compiler for Recursive Deep Learning Models},
 url = {https://proceedings.mlsys.org/paper_files/paper/2021/file/eca986d585a03890a412587a2f5ccb43-Paper.pdf},
 volume = {3},
 year = {2021}
}

@inproceedings{feng:tensorir,
author = {Feng, Siyuan and Hou, Bohan and Jin, Hongyi and Lin, Wuwei and Shao, Junru and Lai, Ruihang and Ye, Zihao and Zheng, Lianmin and Yu, Cody Hao and Yu, Yong and Chen, Tianqi},
title = {TensorIR: An Abstraction for Automatic Tensorized Program Optimization},
year = {2023},
isbn = {9781450399166},
publisher = {Association for Computing Machinery},
address = {New York, NY, USA},
url = {https://doi.org/10.1145/3575693.3576933},
doi = {10.1145/3575693.3576933},
booktitle = {Proceedings of the 28th ACM International Conference on Architectural Support for Programming Languages and Operating Systems, Volume 2},
pages = {804–817},
numpages = {14},
location = {Vancouver, BC, Canada},
series = {ASPLOS 2023}
}

@article{gyurgyik:scion,
author = {Gyurgyik, Christophe and Root, Alexander J and Kjolstad, Fredrik},
title = {Decoupling Data Layouts from Bounding Volume Hierarchies},
year = {2026},
issue_date = {June 2026},
publisher = {Association for Computing Machinery},
address = {New York, NY, USA},
volume = {10},
number = {PLDI},
url = {https://doi.org/10.1145/3808253},
doi = {10.1145/3808253},
journal = {Proc. ACM Program. Lang.},
month = jun,
articleno = {175},
numpages = {26}
}

@inproceedings{hagedorn:fireiron,
author = {Hagedorn, Bastian and Elliott, Archibald Samuel and Barthels, Henrik and Bodik, Rastislav and Grover, Vinod},
title = {Fireiron: A Data-Movement-Aware Scheduling Language for GPUs},
year = {2020},
isbn = {9781450380751},
publisher = {Association for Computing Machinery},
address = {New York, NY, USA},
url = {https://doi.org/10.1145/3410463.3414632},
doi = {10.1145/3410463.3414632},
booktitle = {Proceedings of the ACM International Conference on Parallel Architectures and Compilation Techniques},
pages = {71–82},
numpages = {12},
location = {Virtual Event, GA, USA},
series = {PACT '20}
}

@article{hagedorn:rise-elevate,
author = {Hagedorn, Bastian and Lenfers, Johannes and K{\oe}hler, Thomas and Qin, Xueying and Gorlatch, Sergei and Steuwer, Michel},
title = {Achieving high-performance the functional way: a functional pearl on expressing high-performance optimizations as rewrite strategies},
year = {2020},
issue_date = {August 2020},
publisher = {Association for Computing Machinery},
address = {New York, NY, USA},
volume = {4},
number = {ICFP},
url = {https://doi.org/10.1145/3408974},
doi = {10.1145/3408974},
journal = {Proc. ACM Program. Lang.},
month = aug,
articleno = {92},
numpages = {29}
}

@misc{haj-ali:protuner,
      title={ProTuner: Tuning Programs with Monte Carlo Tree Search}, 
      author={Ameer Haj-Ali and Hasan Genc and Qijing Huang and William Moses and John Wawrzynek and Krste Asanović and Ion Stoica},
      year={2020},
      eprint={2005.13685},
      archivePrefix={arXiv},
      primaryClass={cs.DC},
      url={https://arxiv.org/abs/2005.13685}, 
}

@software{halide:repo,
  author       = {{The Halide Developers}},
  title        = {Halide},
  version      = {21.0.0},
  year         = {2025},
  url          = {https://github.com/halide/Halide},
  lastaccessed = {October 22, 2026},
}

@misc{hall:taxonomy,
      title={Scheduling Languages: A Past, Present, and Future Taxonomy}, 
      author={Mary Hall and Cosmin Oancea and Anne C. Elster and Ari Rasch and Sameeran Joshi and Amir Mohammad Tavakkoli and Richard Schulze},
      year={2024},
      eprint={2410.19927},
      archivePrefix={arXiv},
      primaryClass={cs.PL},
      url={https://arxiv.org/abs/2410.19927}, 
}

@article{hellsten:baco,
  title={BaCO: A Fast and Portable Bayesian Compiler Optimization Framework},
  author={Erik Hellsten and Artur Souza and Johannes Lenfers and Rubens Lacouture and Olivia Hsu and Adel Ejjeh and Fredrik Kjolstad and Michel Steuwer and Kunle Olukotun and Luigi Nardi},
  journal={International Conference on Architectural Support for Programming Languages and Operating Systems},
  year={2023},
  month={April}
}

@misc{hong:autocomp,
      title={Autocomp: A Powerful and Portable Code Optimizer for Tensor Accelerators}, 
      author={Charles Hong and Sahil Bhatia and Alvin Cheung and Yakun Sophia Shao},
      year={2025},
      eprint={2505.18574},
      archivePrefix={arXiv},
      primaryClass={cs.PL},
      url={https://arxiv.org/abs/2505.18574}, 
}

@INPROCEEDINGS{hong:llm-compilation,
  author={Hong, Charles and Bhatia, Sahil and Haan, Altan and Dong, Shengjun Kris and Nikiforov, Dima and Cheung, Alvin and Shao, Yakun Sophia},
  booktitle={2024 IEEE LLM Aided Design Workshop (LAD)}, 
  title={LLM-Aided Compilation for Tensor Accelerators}, 
  year={2024},
  volume={},
  number={},
  pages={1-14},
  doi={10.1109/LAD62341.2024.10691748}}

@inproceedings{hsu:sam,
author = {Hsu, Olivia and Strange, Maxwell and Sharma, Ritvik and Won, Jaeyeon and Olukotun, Kunle and Emer, Joel S. and Horowitz, Mark A. and Kj\o{}lstad, Fredrik},
title = {The Sparse Abstract Machine},
year = {2023},
isbn = {9781450399180},
publisher = {Association for Computing Machinery},
address = {New York, NY, USA},
url = {https://doi.org/10.1145/3582016.3582051},
doi = {10.1145/3582016.3582051},
booktitle = {Proceedings of the 28th ACM International Conference on Architectural Support for Programming Languages and Operating Systems, Volume 3},
pages = {710–726},
numpages = {17},
location = {Vancouver, BC, Canada},
series = {ASPLOS 2023}
}

@inproceedings{hsu:stardust,
author = {Hsu, Olivia and Rucker, Alexander and Zhao, Tian and Desai, Varun and Olukotun, Kunle and Kjolstad, Fredrik},
title = {Stardust: Compiling Sparse Tensor Algebra to a Reconfigurable Dataflow Architecture},
year = {2025},
isbn = {9798400712753},
publisher = {Association for Computing Machinery},
address = {New York, NY, USA},
url = {https://doi.org/10.1145/3696443.3708918},
doi = {10.1145/3696443.3708918},
booktitle = {Proceedings of the 23rd ACM/IEEE International Symposium on Code Generation and Optimization},
pages = {628–643},
numpages = {16},
location = {Las Vegas, NV, USA},
series = {CGO '25}
}

@article{hu:taichi,
author = {Hu, Yuanming and Li, Tzu-Mao and Anderson, Luke and Ragan-Kelley, Jonathan and Durand, Fr{\'e}do},
title = {Taichi: a language for high-performance computation on spatially sparse data structures},
year = {2019},
issue_date = {December 2019},
publisher = {Association for Computing Machinery},
address = {New York, NY, USA},
volume = {38},
number = {6},
issn = {0730-0301},
url = {https://doi.org/10.1145/3355089.3356506},
doi = {10.1145/3355089.3356506},
journal = {ACM Trans. Graph.},
month = nov,
articleno = {201},
numpages = {16}
}

@inproceedings{ikarashi:exo,
author = {Ikarashi, Yuka and Bernstein, Gilbert Louis and Reinking, Alex and Genc, Hasan and Ragan-Kelley, Jonathan},
title = {Exocompilation for productive programming of hardware accelerators},
year = {2022},
isbn = {9781450392655},
publisher = {Association for Computing Machinery},
address = {New York, NY, USA},
url = {https://doi.org/10.1145/3519939.3523446},
doi = {10.1145/3519939.3523446},
booktitle = {Proceedings of the 43rd ACM SIGPLAN International Conference on Programming Language Design and Implementation},
pages = {703–718},
numpages = {16},
location = {San Diego, CA, USA},
series = {PLDI 2022}
}

@inproceedings{ikarashi:exo2,
author = {Ikarashi, Yuka and Qian, Kevin and Droubi, Samir and Reinking, Alex and Bernstein, Gilbert Louis and Ragan-Kelley, Jonathan},
title = {Exo 2: Growing a Scheduling Language},
year = {2025},
isbn = {9798400706981},
publisher = {Association for Computing Machinery},
address = {New York, NY, USA},
url = {https://doi.org/10.1145/3669940.3707218},
doi = {10.1145/3669940.3707218},
booktitle = {Proceedings of the 30th ACM International Conference on Architectural Support for Programming Languages and Operating Systems, Volume 1},
pages = {426–444},
numpages = {19},
location = {Rotterdam, Netherlands},
series = {ASPLOS '25}
}

@inproceedings{ikarashi:roly-poly,
  author={Ikarashi, Yuka and Ragan-Kelley, Jonathan and Fukusato, Tsukasa and Kato, Jun and Igarashi, Takeo},
  booktitle={2021 IEEE Symposium on Visual Languages and Human-Centric Computing (VL/HCC)}, 
  title={Guided Optimization for Image Processing Pipelines}, 
  year={2021},
  volume={},
  number={},
  pages={1-5},
  doi={10.1109/VL/HCC51201.2021.9576341}}

@article{kennedy:parascope,
  author={Kennedy, K. and McKinley, K.S. and Tseng, C.-W.},
  journal={IEEE Transactions on Parallel and Distributed Systems}, 
  title={Interactive parallel programming using the ParaScope Editor}, 
  year={1991},
  volume={2},
  number={3},
  pages={329-341},
  doi={10.1109/71.86108}}

@inproceedings{kjolstad:taco,
  author = {Kjolstad, Fredrik and Kamil, Shoaib and Chou, Stephen and Lugato, David and Amarasinghe, Saman},
  title = {The Tensor Algebra Compiler},
  journal = {Proc. ACM Program. Lang.},
  issue_date = {October 2017},
  volume = {1},
  number = {OOPSLA},
  month = oct,
  year = {2017},
  issn = {2475-1421},
  pages = {77:1--77:29},
  articleno = {77},
  numpages = {29},
  url = {http://doi.acm.org/10.1145/3133901},
  doi = {10.1145/3133901},
  acmid = {3133901},
  publisher = {ACM},
  address = {New York, NY, USA}
}

@inproceedings{kjolstad:taco-sched,
  title={Tensor Algebra Compilation with Workspaces},
  author={Fredrik Kjolstad and Willow Ahrens and Shoaib Kamil and Saman Amarasinghe},
  journal={International Symposium on Code Generation and Optimization},
  year={2019},
  month={February}
}

@inproceedings{koehler:guided-eqsat,
author = {Koehler, Thomas and Goens, Andr\'{e}s and Bhat, Siddharth and Grosser, Tobias and Trinder, Phil and Steuwer, Michel},
title = {Guided Equality Saturation},
year = {2024},
issue_date = {January 2024},
publisher = {Association for Computing Machinery},
address = {New York, NY, USA},
volume = {8},
number = {POPL},
url = {https://doi.org/10.1145/3632900},
doi = {10.1145/3632900},
journal = {Proc. ACM Program. Lang.},
month = jan,
articleno = {58},
numpages = {32}
}

@inproceedings{koeplinger:spatial,
author = {Koeplinger, David and Feldman, Matthew and Prabhakar, Raghu and Zhang, Yaqi and Hadjis, Stefan and Fiszel, Ruben and Zhao, Tian and Nardi, Luigi and Pedram, Ardavan and Kozyrakis, Christos and Olukotun, Kunle},
title = {Spatial: a language and compiler for application accelerators},
year = {2018},
isbn = {9781450356985},
publisher = {Association for Computing Machinery},
address = {New York, NY, USA},
url = {https://doi.org/10.1145/3192366.3192379},
doi = {10.1145/3192366.3192379},
booktitle = {Proceedings of the 39th ACM SIGPLAN Conference on Programming Language Design and Implementation},
pages = {296–311},
numpages = {16},
location = {Philadelphia, PA, USA},
series = {PLDI 2018}
}

@article{koul:onyx,
  title={Onyx: A 12nm 756 GOPS/W Coarse-Grained Reconfigurable Array for Accelerating Dense and Sparse Applications},
  author={Kalhan Koul and Maxwell Strange and Jackson Melchert and Alex Carsello and Yuchen Mei and Olivia Hsu and Taeyoung Kong and Po-Han Chen and Huifeng Ke and Keyi Zhang and Qiaoyi Liu and Gedeon Nyengele and Akhilesh Balasingam and Jayashree Adivarahan and Ritvik Sharma and Zhouhua Xie and Christopher Torng and Joel Emer and Fredrik Kjolstad and Mark Horowitz and Priyanka Raina},
  journal={IEEE Symposium on VLSI Technology and Circuits (VLSI)},
  year={2024},
  month={June}
}

@article{koul:onyx-jssc, 
  author={Koul, Kalhan and Hsu, Olivia and Mei, Yuchen and Gautham Ravipati, Sai and Strange, Maxwell and Melchert, Jackson and Carsello, Alex and Kong, Taeyoung and Chen, Po-Han and Ke, Huifeng and Zhang, Keyi and Liu, Qiaoyi and Nyengele, Gedeon and Xie, Zhouhua and Balasingam, Akhilesh and Adivarahan, Jayashree and Sharma, Ritvik and Torng, Christopher and Emer, Joel S. and Kjolstad, Fredrik and Horowitz, Mark and Raina, Priyanka}, 
  journal={IEEE Journal of Solid-State Circuits}, 
  title={Onyx: A 12-nm Programmable Accelerator for Dense and Sparse Applications}, 
  year={2025}, 
  volume={}, 
  number={}, 
  pages={1-13}, 
  doi={10.1109/JSSC.2025.3604724}}

@inproceedings{lacouture:fuseflow,
author = {Lacouture, Rubens and Zhang, Nathan and Sharma, Ritvik and Siracusa, Marco and Kjolstad, Fredrik and Olukotun, Kunle and Hsu, Olivia},
title = {FuseFlow: A Fusion-Centric Compilation Framework for Sparse Deep Learning on Streaming Dataflow},
year = {2026},
isbn = {9798400723599},
publisher = {Association for Computing Machinery},
address = {New York, NY, USA},
url = {https://doi.org/10.1145/3779212.3790165},
doi = {10.1145/3779212.3790165},
booktitle = {Proceedings of the 31st ACM International Conference on Architectural Support for Programming Languages and Operating Systems, Volume 2},
pages = {798–820},
numpages = {23},
location = {Pittsburgh, PA, USA},
series = {ASPLOS '26}
}

@inproceedings{lacouture:llm-autosparse,
title={{LLM}-Guided Autoscheduling for Large-Scale Sparse Machine Learning},
author={Rubens Lacouture and Genghan Zhang and Konstantin Hossfeld and Tian Zhao and Kunle Olukotun},
booktitle={Machine Learning for Systems 2025},
year={2025},
url={https://openreview.net/forum?id=7H9qWe8lLO}
}

@inproceedings{lai:hetero-cl,
author = {Lai, Yi-Hsiang and Chi, Yuze and Hu, Yuwei and Wang, Jie and Yu, Cody Hao and Zhou, Yuan and Cong, Jason and Zhang, Zhiru},
title = {HeteroCL: A Multi-Paradigm Programming Infrastructure for Software-Defined Reconfigurable Computing},
year = {2019},
isbn = {9781450361378},
publisher = {Association for Computing Machinery},
address = {New York, NY, USA},
url = {https://doi.org/10.1145/3289602.3293910},
doi = {10.1145/3289602.3293910},
booktitle = {Proceedings of the 2019 ACM/SIGDA International Symposium on Field-Programmable Gate Arrays},
pages = {242–251},
numpages = {10},
location = {Seaside, CA, USA},
series = {FPGA '19}
}

@article{li:differentiable-halide,
author = {Li, Tzu-Mao and Gharbi, Micha{\"e}l and Adams, Andrew and Durand, Fr{\'e}do and Ragan-Kelley, Jonathan},
title = {Differentiable programming for image processing and deep learning in halide},
year = {2018},
issue_date = {August 2018},
publisher = {Association for Computing Machinery},
address = {New York, NY, USA},
volume = {37},
number = {4},
issn = {0730-0301},
url = {https://doi.org/10.1145/3197517.3201383},
doi = {10.1145/3197517.3201383},
journal = {ACM Trans. Graph.},
month = jul,
articleno = {139},
numpages = {13}
}

@article{liao:suif-explorer,
author = {Liao, Shih-Wei and Diwan, Amer and Bosch, Robert P. and Ghuloum, Anwar and Lam, Monica S.},
title = {SUIF Explorer: an interactive and interprocedural parallelizer},
year = {1999},
issue_date = {Aug. 1999},
publisher = {Association for Computing Machinery},
address = {New York, NY, USA},
volume = {34},
number = {8},
issn = {0362-1340},
url = {https://doi.org/10.1145/329366.301108},
doi = {10.1145/329366.301108},
journal = {SIGPLAN Not.},
month = may,
pages = {37–48},
numpages = {12}
}

@article{liu:atl-pldi24,
author = {Liu, Amanda and Bernstein, Gilbert and Chlipala, Adam and Ragan-Kelley, Jonathan},
title = {A Verified Compiler for a Functional Tensor Language},
year = {2024},
issue_date = {June 2024},
publisher = {Association for Computing Machinery},
address = {New York, NY, USA},
volume = {8},
number = {PLDI},
url = {https://doi.org/10.1145/3656390},
doi = {10.1145/3656390},
journal = {Proc. ACM Program. Lang.},
month = jun,
articleno = {160},
numpages = {23}
}

@article{liu:atl-pldi26,
author = {Liu, Amanda and Bernstein, Gilbert Louis and Kamil, Shoaib and Chlipala, Adam and Ragan-Kelley, Jonathan},
title = {A Mechanized Algebra of Verified Data Structures for Optimizing Sparse Tensor Programs},
year = {2026},
issue_date = {June 2026},
publisher = {Association for Computing Machinery},
address = {New York, NY, USA},
volume = {10},
number = {PLDI},
url = {https://doi.org/10.1145/3808261},
doi = {10.1145/3808261},
journal = {Proc. ACM Program. Lang.},
month = jun,
articleno = {183},
numpages = {23}
}

@article{liu:atl-popl22,
author = {Liu, Amanda and Bernstein, Gilbert Louis and Chlipala, Adam and Ragan-Kelley, Jonathan},
title = {Verified tensor-program optimization via high-level scheduling rewrites},
year = {2022},
issue_date = {January 2022},
publisher = {Association for Computing Machinery},
address = {New York, NY, USA},
volume = {6},
number = {POPL},
url = {https://doi.org/10.1145/3498717},
doi = {10.1145/3498717},
journal = {Proc. ACM Program. Lang.},
month = jan,
articleno = {55},
numpages = {28}
}

@article{lubin:honeybee,
author = {Lubin, Justin and Ziegler, Parker and Chasins, Sarah E.},
title = {Programming by Navigation},
year = {2025},
issue_date = {June 2025},
publisher = {Association for Computing Machinery},
address = {New York, NY, USA},
volume = {9},
number = {PLDI},
url = {https://doi.org/10.1145/3729264},
doi = {10.1145/3729264},
journal = {Proc. ACM Program. Lang.},
month = jun,
articleno = {165},
numpages = {28}
}

@inproceedings{merouani:compilot,
author = {Merouani, Massinissa and Bernou, Islem Kara and Baghdadi, Riyadh},
title = {Agentic Auto-Scheduling: An Experimental Study of LLM-Guided Loop Optimization},
year = {2025},
publisher = {IEEE Press},
url = {https://doi.org/10.1109/PACT65351.2025.00027},
doi = {10.1109/PACT65351.2025.00027},
booktitle = {Proceedings of the 2025 34th International Conference on Parallel Architectures and Compilation Techniques},
pages = {186–200},
numpages = {15},
series = {PACT '25}
}

@article{mullapudi:halide-auto,
author = {Mullapudi, Ravi Teja and Adams, Andrew and Sharlet, Dillon and Ragan-Kelley, Jonathan and Fatahalian, Kayvon},
title = {Automatically scheduling halide image processing pipelines},
year = {2016},
issue_date = {July 2016},
publisher = {Association for Computing Machinery},
address = {New York, NY, USA},
volume = {35},
number = {4},
issn = {0730-0301},
url = {https://doi.org/10.1145/2897824.2925952},
doi = {10.1145/2897824.2925952},
journal = {ACM Trans. Graph.},
month = jul,
articleno = {83},
numpages = {11}
}

@inproceedings{mullapudi:polymage,
author = {Mullapudi, Ravi Teja and Vasista, Vinay and Bondhugula, Uday},
title = {PolyMage: Automatic Optimization for Image Processing Pipelines},
year = {2015},
isbn = {9781450328357},
publisher = {Association for Computing Machinery},
address = {New York, NY, USA},
url = {https://doi.org/10.1145/2694344.2694364},
doi = {10.1145/2694344.2694364},
booktitle = {Proceedings of the Twentieth International Conference on Architectural Support for Programming Languages and Operating Systems},
pages = {429–443},
numpages = {15},
location = {Istanbul, Turkey},
series = {ASPLOS '15}
}

@misc{nardi:hyper-mapper,
      title={Practical Design Space Exploration}, 
      author={Luigi Nardi and David Koeplinger and Kunle Olukotun},
      year={2019},
      eprint={1810.05236},
      archivePrefix={arXiv},
      primaryClass={cs.LG},
      url={https://arxiv.org/abs/1810.05236}, 
}

@inproceedings{nayak:teaal,
author = {Nayak, Nandeeka and Odemuyiwa, Toluwanimi O. and Ugare, Shubham and Fletcher, Christopher and Pellauer, Michael and Emer, Joel},
title = {TeAAL: A Declarative Framework for Modeling Sparse Tensor Accelerators},
year = {2023},
isbn = {9798400703294},
publisher = {Association for Computing Machinery},
address = {New York, NY, USA},
url = {https://doi.org/10.1145/3613424.3623791},
doi = {10.1145/3613424.3623791},
booktitle = {Proceedings of the 56th Annual IEEE/ACM International Symposium on Microarchitecture},
pages = {1255–1270},
numpages = {16},
location = {Toronto, ON, Canada},
series = {MICRO '23}
}

@inproceedings{ouyang:kernelbench,
title={KernelBench: Can {LLM}s Write Efficient {GPU} Kernels?},
author={Anne Ouyang and Simon Guo and Simran Arora and Alex L Zhang and William Hu and Christopher Re and Azalia Mirhoseini},
booktitle={Forty-second International Conference on Machine Learning},
year={2025},
url={https://openreview.net/forum?id=yeoN1iQT1x}
}

@article{papenhausen:pumav,
  author={Papenhausen, Eric and Langston, M. Harper and Meister, Benoit and Lethin, Richard A. and Mueller, Klaus},
  journal={IEEE Computer Graphics and Applications}, 
  title={PUMA-V: Optimizing Parallel Code Performance Through Interactive Visualization}, 
  year={2019},
  volume={39},
  number={1},
  pages={84-99},
  doi={10.1109/MCG.2018.2877113}}

@inbook{ragan-kelley:decouple,
author = {Ragan-Kelley, Jonathan and Adams, Andrew and Paris, Sylvain and Levoy, Marc and Amarasinghe, Saman and Durand, Fr\'{e}do},
title = {Decoupling Algorithms from Schedules for Easy Optimization of Image Processing Pipelines},
year = {2023},
isbn = {9798400708978},
publisher = {Association for Computing Machinery},
address = {New York, NY, USA},
edition = {1},
url = {https://doi.org/10.1145/3596711.3596751},
booktitle = {Seminal Graphics Papers: Pushing the Boundaries, Volume 2},
articleno = {39},
numpages = {12}
}

@article{ragan-kelley:halide,
author = {Ragan-Kelley, Jonathan and Barnes, Connelly and Adams, Andrew and Paris, Sylvain and Durand, Fr{\'e}do and Amarasinghe, Saman},
title = {Halide: a language and compiler for optimizing parallelism, locality, and recomputation in image processing pipelines},
year = {2013},
issue_date = {June 2013},
publisher = {Association for Computing Machinery},
address = {New York, NY, USA},
volume = {48},
number = {6},
issn = {0362-1340},
url = {https://doi.org/10.1145/2499370.2462176},
doi = {10.1145/2499370.2462176},
journal = {SIGPLAN Not.},
month = jun,
pages = {519–530},
numpages = {12}
}

@inproceedings{ramstad:llm-hls,
author = {Ramstad, Max and Agostini, Nicolas Bohm and Tumeo, Antonino},
title = {LLM-Driven Optimization for High-Level Synthesis},
year = {2026},
isbn = {9798400725685},
publisher = {Association for Computing Machinery},
address = {New York, NY, USA},
url = {https://doi.org/10.1145/3801487.3806068},
doi = {10.1145/3801487.3806068},
booktitle = {Proceedings of the 23rd ACM International Conference on Computing Frontiers},
pages = {323–326},
numpages = {4},
location = {Catania, Italy},
series = {CF '26}
}

@misc{reinking:halide-semantics,
      title={Formal Semantics for the Halide Language}, 
      author={Alex Reinking and Gilbert Louis Bernstein and Jonathan Ragan-Kelley},
      year={2022},
      eprint={2210.15740},
      archivePrefix={arXiv},
      primaryClass={cs.PL},
      url={https://arxiv.org/abs/2210.15740}, 
}

@article{senanayake:sparse-iteration,
  title={A sparse iteration space transformation framework for sparse tensor algebra},
  author={Senanayake, Ryan and Hong, Changwan and Wang, Ziheng and Wilson, Amalee and Chou, Stephen and Kamil, Shoaib and Amarasinghe, Saman and Kjolstad, Fredrik},
  journal={Proceedings of the ACM on Programming Languages},
  volume={4},
  number={OOPSLA},
  pages={1--30},
  year={2020},
  publisher={ACM New York, NY, USA}
}

@inproceedings{shao:metaschedule,
author = {Shao, Junru and Zhou, Xiyou and Feng, Siyuan and Hou, Bohan and Lai, Ruihang and Jin, Hongyi and Lin, Wuwei and Masuda, Masahiro and Yu, Cody Hao and Chen, Tianqi},
title = {Tensor program optimization with probabilistic programs},
year = {2022},
isbn = {9781713871088},
publisher = {Curran Associates Inc.},
address = {Red Hook, NY, USA},
booktitle = {Proceedings of the 36th International Conference on Neural Information Processing Systems},
articleno = {2593},
numpages = {14},
location = {New Orleans, LA, USA},
series = {NIPS '22}
}

@inproceedings{schulze:mdh,
author = {Schulze, Richard and Gorlatch, Sergei and Rasch, Ari},
title = {Reduction-Aware Directive-Based Programming via Multi-Dimensional Homomorphisms},
year = {2025},
isbn = {9798400718717},
publisher = {Association for Computing Machinery},
address = {New York, NY, USA},
url = {https://doi.org/10.1145/3731599.3767572},
doi = {10.1145/3731599.3767572},
booktitle = {Proceedings of the SC '25 Workshops of the International Conference for High Performance Computing, Networking, Storage and Analysis},
pages = {2100–2113},
numpages = {14},
location = {St Louis, MO, USA},
series = {SC Workshops '25}
}

@article{sioutas:reuse,
author = {Sioutas, Savvas and Stuijk, Sander and Waeijen, Luc and Basten, Twan and Corporaal, Henk and Somers, Lou},
title = {Schedule Synthesis for Halide Pipelines through Reuse Analysis},
year = {2019},
issue_date = {June 2019},
publisher = {Association for Computing Machinery},
address = {New York, NY, USA},
volume = {16},
number = {2},
issn = {1544-3566},
url = {https://doi.org/10.1145/3310248},
doi = {10.1145/3310248},
journal = {ACM Trans. Archit. Code Optim.},
month = apr,
articleno = {10},
numpages = {22}
}

@article{sioutas:halide-gpu,
author = {Sioutas, Savvas and Stuijk, Sander and Basten, Twan and Corporaal, Henk and Somers, Lou},
title = {Schedule Synthesis for Halide Pipelines on GPUs},
year = {2020},
issue_date = {September 2020},
publisher = {Association for Computing Machinery},
address = {New York, NY, USA},
volume = {17},
number = {3},
issn = {1544-3566},
url = {https://doi.org/10.1145/3406117},
doi = {10.1145/3406117},
journal = {ACM Trans. Archit. Code Optim.},
month = aug,
articleno = {23},
numpages = {25}
}

@inproceedings{solar-lezama:sketching,
author = {Solar-Lezama, Armando},
title = {The Sketching Approach to Program Synthesis},
year = {2009},
isbn = {9783642106712},
publisher = {Springer-Verlag},
address = {Berlin, Heidelberg},
url = {https://doi.org/10.1007/978-3-642-10672-9_3},
doi = {10.1007/978-3-642-10672-9_3},
booktitle = {Proceedings of the 7th Asian Symposium on Programming Languages and Systems},
pages = {4–13},
numpages = {10},
location = {Seoul, Korea},
series = {APLAS '09}
}

@inproceedings{steuwer:lift,
  author={Steuwer, Michel and Remmelg, Toomas and Dubach, Christophe},
  booktitle={2017 IEEE/ACM International Symposium on Code Generation and Optimization (CGO)}, 
  title={LIFT: A functional data-parallel IR for high-performance GPU code generation}, 
  year={2017},
  volume={},
  number={},
  pages={74-85},
  doi={10.1109/CGO.2017.7863730}}

@article{sundram:recuma,
  title={Compiling Recurrences over Dense and Sparse Arrays},
  author={Sundram, Shiv and Tariq, Muhammad Usman and Kjolstad, Fredrik},
  journal={Proceedings of the ACM on Programming Languages},
  volume={8},
  number={OOPSLA1},
  pages={250--275},
  year={2024},
  publisher={ACM, New York, NY, USA}
}

@inproceedings{swann:seer,
  title={Seer: Predictive runtime kernel selection for irregular problems},
  author={Swann, Ryan and Osama, Muhammad and Sangaiah, Karthik and Mahmud, Jalal},
  booktitle={2024 IEEE/ACM International Symposium on Code Generation and Optimization (CGO)},
  pages={133--142},
  year={2024},
  organization={IEEE}
}

@article{tariq:reptile,
  title={REPTILE: Performant Tiling of Recurrences},
  author={Tariq, Muhammad Usman and Sundram, Shiv and Kjolstad, Fredrik},
  journal={Proceedings of the ACM on Programming Languages},
  volume={9},
  number={OOPSLA2},
  pages={670--696},
  year={2025},
  publisher={ACM, New York, NY, USA}
}

@article{venkat:loop,
  title={Loop and data transformations for sparse matrix code},
  author={Venkat, Anand and Hall, Mary and Strout, Michelle},
  journal={ACM SIGPLAN Notices},
  volume={50},
  number={6},
  pages={521--532},
  year={2015},
  publisher={ACM New York, NY, USA}
}

@article{venkat:swirl,
author = {Venkat, Anand and Rusira, Tharindu and Barik, Raj and Hall, Mary and Truong, Leonard},
editor = {Jack Dongarra and Bernard Tourancheau},
title = {SWIRL: High-performance many-core CPU code generation for deep neural networks},
year = {2019},
issue_date = {Nov 2019},
publisher = {Sage Publications, Inc.},
address = {USA},
volume = {33},
number = {6},
issn = {1094-3420},
url = {https://doi.org/10.1177/1094342019866247},
doi = {10.1177/1094342019866247},
journal = {Int. J. High Perform. Comput. Appl.},
month = nov,
pages = {1275–1289},
numpages = {15}
}

@inproceedings{willsey:egg,
author = {Willsey, Max and Nandi, Chandrakana and Wang, Yisu Remy and Flatt, Oliver and Tatlock, Zachary and Panchekha, Pavel},
title = {egg: Fast and extensible equality saturation},
year = {2021},
issue_date = {January 2021},
publisher = {Association for Computing Machinery},
address = {New York, NY, USA},
volume = {5},
number = {POPL},
url = {https://doi.org/10.1145/3434304},
doi = {10.1145/3434304},
journal = {Proc. ACM Program. Lang.},
month = jan,
articleno = {23},
numpages = {29}
}

@inproceedings{won:waco,
  title={Waco: learning workload-aware co-optimization of the format and schedule of a sparse tensor program},
  author={Won, Jaeyeon and Mendis, Charith and Emer, Joel S and Amarasinghe, Saman},
  booktitle={Proceedings of the 28th ACM International Conference on Architectural Support for Programming Languages and Operating Systems, Volume 2},
  pages={920--934},
  year={2023}
}

@inproceedings{yadav:distal,
  title={DISTAL: the distributed tensor algebra compiler},
  author={Yadav, Rohan and Aiken, Alex and Kjolstad, Fredrik},
  booktitle={Proceedings of the 43rd ACM SIGPLAN International Conference on Programming Language Design and Implementation},
  pages={286--300},
  year={2022}
}

@inproceedings{yadav:spdistal,
  title={SpDISTAL: compiling distributed sparse tensor computations},
  author={Yadav, Rohan and Aiken, Alex and Kjolstad, Fredrik},
  booktitle={SC22: International Conference for High Performance Computing, Networking, Storage and Analysis},
  pages={1--15},
  year={2022},
  organization={IEEE}
}

@article{yan:scorch,
  title={Fast Autoscheduling for Sparse ML Frameworks},
  author={Bobby Yan and Alexander J Root and Trevor Gale and David Broman and Fredrik Kjolstad},
  journal={International Symposium on Code Generation and Optimization},
  year={2026},
  month={February}
}

@inproceedings{zhang:dam,
  author={Zhang, Nathan and Lacouture, Rubens and Sohn, Gina and Mure, Paul and Zhang, Qizheng and Kjolstad, Fredrik and Olukotun, Kunle},
  booktitle={2024 ACM/IEEE 51st Annual International Symposium on Computer Architecture (ISCA)}, 
  title={The Dataflow Abstract Machine Simulator Framework}, 
  year={2024},
  volume={},
  number={},
  pages={532-547},
  doi={10.1109/ISCA59077.2024.00046}
}

@article{zhang:graphit,
author = {Zhang, Yunming and Yang, Mengjiao and Baghdadi, Riyadh and Kamil, Shoaib and Shun, Julian and Amarasinghe, Saman},
title = {GraphIt: a high-performance graph DSL},
year = {2018},
issue_date = {November 2018},
publisher = {Association for Computing Machinery},
address = {New York, NY, USA},
volume = {2},
number = {OOPSLA},
url = {https://doi.org/10.1145/3276491},
doi = {10.1145/3276491},
journal = {Proc. ACM Program. Lang.},
month = oct,
articleno = {121},
numpages = {30}
}

@misc{zhang:ptxbench,
      title={PTXBench: Benchmarking and Adapting LLMs for GPU Kernel Optimization with Architecture-specific PTX}, 
      author={Genghan Zhang and Yixin Dong and Chengze Fan and Zhichen Zeng and Yueming Yuan and Shaowei Zhu and Kunle Olukotun},
      year={2026},
      eprint={2608.17379},
      archivePrefix={arXiv},
      primaryClass={cs.CL},
      url={https://arxiv.org/abs/2608.17379}, 
}

@article{zhao:nautilus,
  title={Nautilus: An Auto-Scheduling Tensor Compiler for Efficient Tiled GPU Kernels},
  author={Zhao, Yifan and Yang, Yuchen and Budiu, Matei and Misailovic, Sasa},
  journal={arXiv preprint arXiv:2604.14825},
  year={2026}
}

@inproceedings{zheng:ansor,
author = {Zheng, Lianmin and Jia, Chengfan and Sun, Minmin and Wu, Zhao and Yu, Cody Hao and Haj-Ali, Ameer and Wang, Yida and Yang, Jun and Zhuo, Danyang and Sen, Koushik and Gonzalez, Joseph E. and Stoica, Ion},
title = {Ansor: generating high-performance tensor programs for deep learning},
year = {2020},
isbn = {978-1-939133-19-9},
publisher = {USENIX Association},
address = {USA},
booktitle = {Proceedings of the 14th USENIX Conference on Operating Systems Design and Implementation},
articleno = {49},
numpages = {17},
series = {OSDI'20}
}

@inproceedings{zheng:flextensor,
author = {Zheng, Size and Liang, Yun and Wang, Shuo and Chen, Renze and Sheng, Kaiwen},
title = {FlexTensor: An Automatic Schedule Exploration and Optimization Framework for Tensor Computation on Heterogeneous System},
year = {2020},
isbn = {9781450371025},
publisher = {Association for Computing Machinery},
address = {New York, NY, USA},
url = {https://doi.org/10.1145/3373376.3378508},
doi = {10.1145/3373376.3378508},
booktitle = {Proceedings of the Twenty-Fifth International Conference on Architectural Support for Programming Languages and Operating Systems},
pages = {859–873},
numpages = {15},
location = {Lausanne, Switzerland},
series = {ASPLOS '20}
}

@inproceedings{zheng:amos,
author = {Zheng, Size and Chen, Renze and Wei, Anjiang and Jin, Yicheng and Han, Qin and Lu, Liqiang and Wu, Bingyang and Li, Xiuhong and Yan, Shengen and Liang, Yun},
title = {AMOS: enabling automatic mapping for tensor computations on spatial accelerators with hardware abstraction},
year = {2022},
isbn = {9781450386104},
publisher = {Association for Computing Machinery},
address = {New York, NY, USA},
url = {https://doi.org/10.1145/3470496.3527440},
doi = {10.1145/3470496.3527440},
booktitle = {Proceedings of the 49th Annual International Symposium on Computer Architecture},
pages = {874–887},
numpages = {14},
location = {New York, New York},
series = {ISCA '22}
}

@article{zinenko:clint,
author = {Zinenko, Oleksandr and Huot, St{\'e}phane and Bastoul, C{\'e}dric},
title = {Visual Program Manipulation in the Polyhedral Model},
year = {2018},
issue_date = {March 2018},
publisher = {Association for Computing Machinery},
address = {New York, NY, USA},
volume = {15},
number = {1},
issn = {1544-3566},
url = {https://doi.org/10.1145/3177961},
doi = {10.1145/3177961},
journal = {ACM Trans. Archit. Code Optim.},
month = mar,
articleno = {16},
numpages = {25}
}

\newpage

\appendix

\section{Benchmarks}
\label{app:benchmarks}
\Cref{tab:benchmarks} lists the programs used in Section~\ref{sec:eval}.
The Halide applications are taken from the generators in the Halide repository~\cite{halide:repo}. 
We use the same applications used to evaluate the Adams2019 autoscheduler~\cite{adams:learning} and refer readers to these sources for descriptions of each application.
We report the number of Halide functions in each pipeline and its input and output resolution.
The FuseFlow benchmarks are the end-to-end sparse models from the FuseFlow evaluation~\cite{lacouture:fuseflow}, which describes them in detail. 
We run each model on the listed datasets. 

\begin{table}
  \centering
  \caption{Benchmarks. 
  On top are the Halide applications from the Halide repository~\cite{halide:repo} used to evaluate \citet{adams:learning}, ordered by the number of Halide functions in the pipeline, with their input/output resolutions. 
  On the bottom are the FuseFlow sparse models from \citet{lacouture:fuseflow} and their datasets and configurations.}
  \label{tab:benchmarks}
  \small
  \begin{tabular}{@{}l r l@{}}
    \toprule
    \multicolumn{3}{@{}l}{\textbf{Halide}} \\
    \textbf{Benchmark} & \textbf{Funcs} & \textbf{Size} \\
    \midrule
    \texttt{matmul}           & 4   & $1024{\times}1024{\times}1024$ \\
    \texttt{conv\_layer}      & 5   & $100{\times}80$, 128\,ch, batch 5 \\
    \texttt{max\_filter}      & 6   & $1536{\times}2560{\times}3$ \\
    \texttt{iir\_blur}        & 6   & $1536{\times}2560{\times}3$ \\
    \texttt{unsharp}          & 9   & $1536{\times}2560{\times}3$ \\
    \texttt{bilateral\_grid}  & 9   & $1536{\times}2560$ \\
    \texttt{hist}             & 10  & $1536{\times}2560{\times}3$ \\
    \texttt{harris}           & 13  & $1536{\times}2560{\times}3$ \\
    \texttt{nl\_means}        & 13  & $512{\times}512{\times}3$ \\
    \texttt{bgu}              & 21  & $192{\times}320{\times}3 \rightarrow 1536{\times}2560{\times}3$ \\
    \texttt{camera\_pipe}     & 34  & $2592{\times}1968 \rightarrow 2560{\times}1920{\times}3$ \\
    \texttt{stencil\_chain}   & 35  & $1536{\times}2560$ \\
    \texttt{interpolate}      & 52  & $1536{\times}2560{\times}3$ \\
    \texttt{lens\_blur}       & 69  & $192{\times}320{\times}3$ (stereo pair) \\
    \texttt{local\_laplacian} & 104 & $1536{\times}2560{\times}3$ \\
    \texttt{resnet\_50}       & 268 & $224{\times}224{\times}3 \rightarrow 1000$ \\
    \midrule
    \multicolumn{3}{@{}l}{\textbf{FuseFlow}} \\
    \textbf{Benchmark} & \multicolumn{2}{l}{\textbf{Datasets / configurations}} \\
    \midrule
    SAE         & \multicolumn{2}{l}{ImageNet, NIH-CXR, LUNA16} \\
    GCN         & \multicolumn{2}{l}{Cora, Cora-ML, OGB-Collab} \\
    GraphSAGE   & \multicolumn{2}{l}{Cora, Cora-ML, OGB-Collab} \\
    GPT-3 Small & \multicolumn{2}{l}{block sizes 32, 64 (BigBird attention)} \\
    \bottomrule
  \end{tabular}
\end{table}

\section{RQ3 Full Results for FuseFlow}
\begin{figure*}[t]
  \centering
  \includegraphics[width=\textwidth]{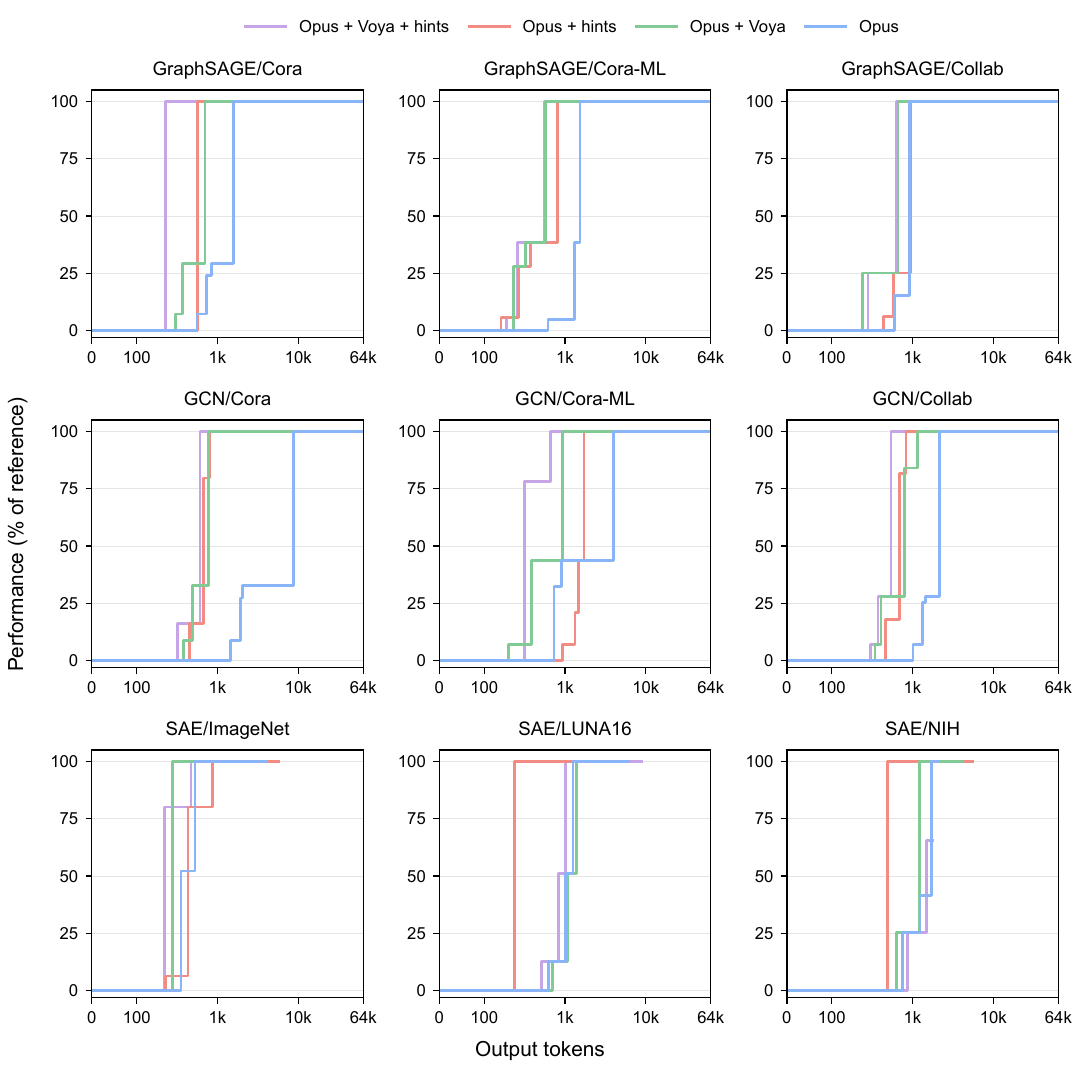}
  \caption{
  FuseFlow schedule performance versus output tokens. 
  Curves show best performance per trial as a percentage of the performance of the best schedule generated by the two approaches. 
  Horizontal axis uses symlog scale. 
  For FuseFlow, agents with \name{} require fewer tokens to reach 100\% performance on average. 
    }
  \label{fig:rq3-agents-fuseflow-app}
\end{figure*}

We evaluate nine of the eleven workload--input configurations from \Cref{sec:eval-rq1}.
\Cref{fig:rq3-agents-fuseflow-app} shows results for all nine configurations.

\label{app:rq3-full}

\end{document}